\documentclass[prc,superscriptaddress,noshowpacs,unsortedaddress,twocolumn,showpacs,preprintnumbers,amsmath,amssymb]{revtex4-1}

\usepackage[dvipdfmx]{graphicx}
\usepackage{amsmath,amssymb,times}
\usepackage{color}
\usepackage{ulem}
\usepackage{bm}
\usepackage{here}

\def\lsim{~\,\makebox(1,1){$\stackrel{<}{\widetilde{}}$}\,~}

\newcommand{\beq}{\begin{equation}}
\newcommand{\eeq}{\end{equation}}
\newcommand{\bea}{\begin{eqnarray}}
\newcommand{\eea}{\end{eqnarray}}

\newcommand{\bfi}[1]{\mbox{\boldmath $#1$}}

\newcommand{\vK}{{\bfi K}}

\newcommand{\vs}{{\bfi s}}

\newcommand{\vrr}{{\bfi r}}
\newcommand{\vR}{{\bfi R}}

\def\a{\alpha}

\begin{document}

\title{Neutron skin in $^{48}$Ca determined from p+$^{48}$Ca and $^{48}$Ca+$^{12}$C scattering
}

\author{Shingo~Tagami}
\affiliation{Department of Physics, Kyushu University, Fukuoka 819-0395, Japan}

\author{Tomotsugu~Wakasa}
\affiliation{Department of Physics, Kyushu University, Fukuoka 819-0395, Japan}

\author{Maya~Takechi}
\affiliation{Niigata University, Niigata 950-2181, Japan}

\author{Jun~Matsui}
\affiliation{Department of Physics, Kyushu University, Fukuoka 819-0395, Japan}

\author{Masanobu~Yahiro}
\email[]{orion093g@gmail.com}
\affiliation{Department of Physics, Kyushu University, Fukuoka 819-0395, Japan}

\begin{abstract}
\begin{description}
\item[Background]
In our previous paper, we determined $r_{\rm skin}^{208}({\rm exp})=0.278 \pm 0.035$~fm from measured 
reaction cross sections $\sigma_{\rm R}$ for p+$^{208}$Pb scattering, using the Kyushu (chiral) $g$-matrix folding model with the densities calculated with the Gogny-D1S Hartree-Fock-Bogoliubov (D1S-GHFB) with 
the angular momentum projection (AMP).   The value agrees with that of PREX2.  
Reaction cross sections $\sigma_{\rm R}$ are available for p+$^{48}$Ca scattering, whereas 
interaction cross sections $\sigma_{\rm I} (\approx\sigma_{\rm R})$ are available 
for $^{48}$Ca + $^{12}$C scattering.  
As for $^{48}$Ca, the high-resolution $E1$ polarizability experiment ($E1$pE) 
yields $r_{\rm skin}^{48}(E1{\rm pE}) =0.14 \sim 0.20~{\rm fm}$.
\item[Purpose]
We  determine $r_{\rm skin}^{48}({\rm exp})$ from the data on $\sigma_{\rm R}$ for 
p+$^{48}$Ca scattering and from the data on $\sigma_{\rm I}$ for $^{48}$Ca+$^{12}$C scattering.
\item[Methods]
We use the Kyushu $g$-matrix folding model with the densities calculated with 
the D1M-GHFB+AMP densities. The D1M-GHFB+AMP proton and 
neutron densities are scaled so as to reproduce the data 
under the condition that the radius $r_{\rm p}$ of the scaled 
proton density equals the data $r_{\rm p}({\rm exp})$ determined from the electron scattering. 
We deduce skin values $r_{\rm skin}=r_{\rm n}({\rm exp})-r_{\rm p}({\rm exp})$ from the resulting 
$r_{\rm n}({\rm exp})$ and the $r_{\rm p}({\rm exp})$ determined from electron scattering. 
The same procedure is taken for D1S-GHFB+AMP. 
\item[Results]
We regard $r_{\rm skin}^{48}(E1{\rm pE})$ as a reference skin value. 
Using the reference skin value and taking D1M-GHFB+AMP, we determine 
$r_{\rm skin}^{48}({\rm exp})=0.158 \pm 0.025$~fm for p+$^{48}$Ca scattering and 
$0.160 \pm 0.058$~fm for  $^{48}$Ca + $^{12}$C scattering. 
\item[Conclusion]
We take the weighted mean and its error for the two skin values. 
The result is   
$r_{\rm skin}^{48}({\rm exp})=0.158 \pm	(0.023)_{\rm exp} \pm (0.012)_{\rm th}~{\rm fm}$.
\end{description}
 \end{abstract}

\maketitle

\section{Introduction and Conclusion}
\label{Sec:Introduction}

{\it Background on experiments:} 

Horowitz, Pollock and Souder proposed a direct measurement 
for neutron skin thickness $r_{\rm skin}=r_{\rm n}-r_{\rm p}$~\cite{PRC.63.025501}, 
where $r_{\rm p}$ and $r_{\rm n}$ are proton and neutron radii, respectively.  
The measurement consists of parity-violating and elastic electron scattering. 
In fact, 
the PREX collaboration has reported a new value, 
\begin{equation}
r_{\rm skin}^{208}({\rm PREX2}) = 0.283\pm 0.071\,{\rm fm}, 
\label{Eq:Experimental constraint 208}
\end{equation}
combining the original Lead Radius EXperiment (PREX)  result \cite{PRL.108.112502,PRC.85.032501} 
with the updated PREX2 result \cite{Adhikari:2021phr}. 
The  value  is most reliable for $r_{\rm skin}^{208}$. 
For $^{48}$Ca, the CREX is still ongoing at Jefferson Lab~\cite{PREX05}. 

As an indirect measurement on $r_{\rm skin}$, the high-resolution $E1$ polarizability 
experiment ($E1$pE)  was made 
for $^{208}$Pb~\cite{Tamii:2011pv} and $^{48}$Ca~\cite{Birkhan:2016qkr} in RCNP. 
The results are
\bea
r_{\rm skin}^{208}(E1{\rm pE}) &=&0.156^{+0.025}_{-0.021}=0.135 \sim 0.181~{\rm fm}, 
\label{Eq:skin-Pb208-E1}
\\
r_{\rm skin}^{48}(E1{\rm pE}) &=&0.14 \sim 0.20~{\rm fm}.  
\label{Eq:skin-Ca48-E1}
\eea

Reaction cross section $\sigma_{\rm R}$ is a standard observable
to determine the matter radius $r_{\rm m}$ and the skin value $r_{\rm skin}$. 
The data $\sigma_{\rm R}({\rm exp})$ are available for $p$+$^{48}$Ca scattering in 
incident energies of $E_{\rm in}=23 \sim 48$~Mev~\cite{Carlson:1994fq}. 
Interaction cross sections $\sigma_{\rm I} (\approx \sigma_{\rm R})$ are available 
for $^{42-51}$Ca + $^{12}$C scattering at 280~MeV per nucleon~\cite{Tanaka:2019pdo}. 

{\it Background on theories:}

 The $r_{\rm skin}^{208}({\rm PREX2}) = 0.283\pm 0.071\,{\rm fm}$ value is most reliable, 
 and provides crucial tests for the equation of state (EoS) 
of nuclear matter 
\cite{PRC.102.051303,AJ.891.148,AP.411.167992,EPJA.56.63,JPG.46.093003}
as well as nuclear structure models.
 For example, Reed {\it et al.} \cite{Reed:2021nqk} 
report a value of the slope parameter $L$ 
and examine the impact of such a stiff symmetry energy 
on some critical neutron-star observables. 
They deduce 
\bea
L = 106 \pm 37=69	\sim 143 ~{\rm MeV}
\eea
from  $r_{\rm skin}^{208}({\rm PREX2})$. 
 It should be noted that the  $r_{\rm skin}^{208}({\rm PREX2}) = 0.283\pm 0.071\,{\rm fm}$ 
is considerably larger than other experimental 
values that are significantly model dependent 
\cite{PRL.87.082501,PRC.82.044611,PRL.107.062502,%
PRL.112.242502}.
 As an exceptional case, a nonlocal dispersive-optical-model 
(DOM) analysis of ${}^{208}{\rm Pb}$ deduces 
$r_{\rm skin}^{\rm DOM} =0.25 \pm 0.05$ fm \cite{PRC.101.044303}.

As an {\it ab initio} method for Ca isotopes,
we should consider the coupled-cluster  (CC) method~\cite{Hagen:2013nca,Hagen:2015yea} with chiral interaction. 
The CC result
$r_{\rm skin}^{48}({\rm CC})=0.12  \sim 0.15~ {\rm fm}$~\cite{Hagen:2015yea}  
is consistent with $r_{\rm skin}^{48}(E1{\rm pE})$. 

Kohno calculated the $g$ matrix  for the symmetric nuclear matter, 
using the Brueckner-Hartree-Fock method with chiral N$^{3}$LO 2NFs and NNLO 3NFs~\cite{Kohno:2012vj}. 
He set $c_D=-2.5$ and $c_E=0.25$ so that  the energy per nucleon can  become minimum 
at $\rho = \rho_{0}$. 
Toyokawa {\it et al.} localized the non-local chiral  $g$ matrix into three-range Gaussian forms~\cite{Toyokawa:2017pdd}, using the localization method proposed 
by the Melbourne group~\cite{von-Geramb-1991,Amos-1994}. 
The resulting local  $g$ matrix is referred to as  ``Kyushu  $g$-matrix''. 
The Kyushu $g$-matrix~\cite{Toyokawa:2017pdd}  is constructed from the chiral nucleon-nucleon (NN) interaction with the cutoff  550~MeV.

In Ref.~\cite{Tagami:2019svt}, 
we tested  the Kyushu $g$-matrix folding model~\cite{Toyokawa:2014yma,Toyokawa:2015zxa,Toyokawa:2017pdd}  
for $^{12}$C+$^{12}$C scattering by comparing measured $\sigma_{\rm R}$ 
with the results of the Kyushu folding model, 
and found that the model is reliable 
in $30  \lsim E_{\rm in} \lsim 100 $~MeV and $250  \lsim E_{\rm in} \lsim 400 $~MeV. 
As for $^{42-51}$Ca+$^{12}$C scattering at $E_{\rm in}=280$~MeV per nucleon scattering, 
we predicted  $\sigma_{\rm R}$, using the Kyushu $g$-matrix folding model 
with the densities calculated with Gogny-D1S HFB (D1S-GHFB) with and without 
the angular momentum projection (AMP). 
Our method is much better than the optical limit of the Glauber model with the Wood-Saxon density. 

There is no overlap between $r_{\rm skin}^{208}({\rm PREX2})$ and  $r_{\rm skin}^{208}(E1{\rm pE})$ 
in one  $\sigma$ level. 
However, we determined a value of $r_{\rm skin}^{208}({\rm exp})$ 
from $\sigma_{\rm R}({\rm exp})$ on  
p+$^{208}$Pb scattering in a range of incident energies, $30 \lsim E_{\rm in} \lsim 100$~MeV~\cite{Tagami:2020bee}, using the Kyushu $g$-matrix folding model;
 the value is $r_{\rm skin}^{208}({\rm exp})=0.278 \pm 0.035$~fm. 
 Our result agrees with $r_{\rm skin}^{208}({\rm PREX2})$.

{\it Background on EoSs:}

Many theoretical  predictions on the symmetry energy $S_{\rm sym}(\rho)$ have been made so far  
by taking several experimental and observational constraints on $S_{\rm sym}(\rho)$ and their combinations. 
In neutron star (NS), the $S_{\rm sym}(\rho)$ and its density ($\rho$) dependence influence strongly the nature within the star.
The symmetry energy $S_{\rm sym}(\rho)$ cannot be measured by experiment directly. 
In place of  $S_{\rm sym}(\rho)$, the neutron-skin thickness $r_{\rm skin}$ is measured to determine 
the slope parameter $L$, since a strong correlation between $r_{\rm skin}^{208}$ and $L$ is well known~\cite{RocaMaza:2011pm}.

We  first accumulate the 204 EoSs from Refs.~\cite{Akmal:1998cf,RocaMaza:2011pm,Ishizuka:2014jsa,Gonzalez-Boquera:2017rzy,D1P-1999,Gonzalez-Boquera:2017uep,Oertel:2016bki,Piekarewicz:2007dx,Lim:2013tqa,Sellahewa:2014nia,Inakura:2015cla,Fattoyev:2013yaa,Steiner:2004fi,Centelles:2010qh,Dutra:2012mb,Brown:2013pwa,Brown:2000pd,Reinhard:2016sce,Tsang:2019ymt,Ducoin:2010as,Fortin:2016hny,Chen:2010qx,Zhao:2016ujh,Zhang:2017hvh,Wang:2014rva,Lourenco:2020qft} 
 in which $r_{\rm skin}^{208}$ and/or $L$ is presented, since 
a strong correlation between $r_{\rm skin}^{208}$ and $L$ is shown. 
In the 204 EoSs of Table I, the number of Gogny EoSs is much smaller than that of Skyrme EoSs.
We then construct two EoSs so that D1M*~\cite{Gonzalez-Boquera:2017rzy} and 
D1P~\cite{Sellahewa:2014nia} may become harder; the two EoSs are referred to 
as D1MK and D1PK, respectively; see the parameter sets of D1MK and D1PK for Table II. 
Eventually, we get the 206 EoSs, as shown in Table I.  
The correlation is more reliable when the number of EoSs is larger. 
For this reason,  we  take  the 206 EoSs.

For the 206 EoSs, both  $r_{\rm skin}^{208}$and $L$ are obtained self-consistently; 
the starting $r_{\rm skin}^{208}$-$L$ relation is  determined from the EoSs in which 
both $r_{\rm skin}^{208}$ and $L$ are presented. 
The resulting relation  
\bea
L=620.39~r_{\rm skin}^{208}-57.963
\label{Eq:skin-L}
\eea
has a strong correlation, because of correlation coefficient $R=0.99$. 
The relation~\eqref{Eq:skin-L} allows us to deduce a constraint on $L$ from 
the PREX2 value of Eq.~\eqref{Eq:Experimental constraint 208}. 
The  range of $L$ are $L = 76 \sim 165$ MeV and $L = 76 \sim 172$ MeV~\cite{RocaMaza:2011pm}.
These values and the value of Ref.~\cite{Reed:2021nqk}  support stiffer EoSs. 
As a famous EoS, we can consider APR~\cite{Akmal:1998cf}. It yields 
$L=57.6$~MeV~\cite{Ishizuka:2014jsa}. The EoS is ruled out.
This is a big problem to be solved, since this calculation is believed to be best for symmetric and neutron matter.  Meanwhile, 
stiffer EoSs allow us to consider the phase transition such as QCD transition in NS.

{\it Purpose:}
We determine a value of $r_{\rm skin}^{48}$ from the experimental data on $\sigma_{\rm R}$ 
for p+$^{48}$Ca scattering in $E_{\rm in}=30 \sim 48$~MeV and the data on 
$\sigma_{\rm I}$ for  $^{48}$Ca+$^{12}$C scattering at $E_{\rm in}=280$~MeV per nucleon scattering.

{\it Methods:}
We use the Kyushu $g$-matrix folding model with the densities calculated with 
D1M-GHFB+AMP, and scale the D1M-GHFB+AMP proton and 
neutron densities so as to reproduce the  experimental data on $\sigma_{\rm R}$ for p+$^{48}$Ca scattering and 
$\sigma_{\rm I} (\approx \sigma_{\rm R})$ for  $^{48}$Ca+$^{12}$C scattering 
under the condition that the radius $r_{\rm p}$ of the scaled 
proton density equals the experimental value $r_{\rm p}({\rm exp})$~\cite{ADNDT.99.69} deduced from the electron scattering. 
The resulting skin value is referred to as $r_{\rm skin}^{48}({\rm exp})$ that has an error coming from 
experimental errors.

D1M~\cite{Goriely:2009zz,Robledo:2018cdj} is an improved version of D1S. 
For comparison, we use D1S in addition to D1M. 
The difference between $r_{\rm skin}^{48}({\rm exp})=0.158 \pm 0.025~{\rm fm}$ 
for D1M and $r_{\rm skin}^{48}({\rm exp})=0.125 \pm	0.02~{\rm fm}$ for D1S is large, and 
D1M is better than D1S for total energy, $\sigma_{\rm R}({\rm exp})$  on 
p+$^{48}$Ca scattering and $\sigma_{\rm I}({\rm exp})$ on $^{48}$Ca+$^{12}$C scattering. 
We then take the result of D1M-GHFB+AMP, 
as shown in Figs.~\ref{fig:N dependence of TE}, \ref{fig:48Ca+12C.pdf}, \ref{fig:p+48Ca-0.pdf}. Derivation of $r_{\rm skin}^{48}({\rm exp})=0.158 \pm 0.025~{\rm fm}$ for D1M 
and $r_{\rm skin}^{48}({\rm exp})=0.125 \pm	0.02~{\rm fm}$ are shown below.

{\it Results:}

Our results for D1M are  
\bea
r_{\rm skin}^{48}({\rm exp})=0.160 \pm 0.058~{\rm fm}=0.102 \sim 0.218~{\rm fm}~~~~
\label{eq-skin-48-D1M}
\eea
for $^{48}$Ca+$^{12}$C scattering at $E_{\rm in}=280$~MeV per nucleon scattering and 
\bea
r_{\rm skin}^{48}({\rm exp})=0.158 \pm 0.025~{\rm fm}=0.134 \sim 0.183~{\rm fm}~~~~
\label{eq-skin-48-D1M}
\eea
for p+$^{48}$Ca scattering in $E_{\rm in}=30 \sim 48$~MeV. 
As for p+$^{48}$Ca scattering, we have taken the upper limit of the data~\cite{Carlson:1994fq}. 
The reason is  that the $r_{\rm skin}^{48}({\rm exp})=0.134 \sim 0.183~{\rm fm}$ 
deduced from the upper limit 
is much closer to $r_{\rm skin}^{48}(E1{\rm pE}) =0.14 \sim 0.20~{\rm fm}$
than $r_{\rm skin}^{48}({\rm exp})=0.044 \sim 0.093~{\rm fm}$ 
from the central value of the data~\cite{Carlson:1994fq}.  
The $r_{\rm skin}^{48}({\rm exp})=0.044 \sim 0.093~{\rm fm}$ is smaller than the lower bound of 
$r_{\rm skin}^{48}=0.125$~fm, as shown in 
Sec.~\ref{the lower bound}.

For comparison, the same procedure is taken for D1S. 
Our results are  
\bea
r_{\rm skin}^{48}({\rm exp})=0.107 \pm	0.059~{\rm fm}~~~~
\label{eq-skin-48-D1S}
\eea
for $^{48}$Ca+$^{12}$C scattering at $E_{\rm in}=280$~MeV per nucleon scattering and 
\bea
r_{\rm skin}^{48}({\rm exp})=0.127 \pm	0.0219~{\rm fm}~~~~
\label{eq-skin-48-D1S}
\eea
for p+$^{48}$Ca scattering in $E_{\rm in}=30 \sim 48$~MeV. 
We take the weighted mean and its error for the two skin values. The result for D1S is 
\bea
r_{\rm skin}^{48}({\rm exp})=0.125 \pm	0.02~{\rm fm}
\eea
The difference between $r_{\rm skin}^{48}({\rm exp})=0.158 \pm 0.025~{\rm fm}$ for D1M and 
$r_{\rm skin}^{48}({\rm exp})=0.125 \pm	0.02~{\rm fm}$ for D1S is large, and 
D1M is  better than D1S for total energy, $\sigma_{\rm R}({\rm exp})$  on 
p+$^{48}$Ca scattering, and $\sigma_{\rm I}({\rm exp})$ on $^{48}$Ca+$^{12}$C scattering. 
Eventually, we take the result of D1M-GHFB+AMP.

{\it Conclusion:}
Finally, our final skin value is 
\bea
r_{\rm skin}^{48}({\rm exp})=0.158 \pm	(0.023)_{\rm exp} \pm (0.012)_{\rm th}~{\rm fm}. 
\eea
The second error is a theoretical error from D1M and D1S.

\section{Framework}

Our framework is Kyushu $g$-matrix folding model calculated with GHFB+AMP.

\subsection{Kyushu $g$-matrix folding model}
\label{Sec:Folding model}

The  Kyushu $g$-matrix folding model is successful in reproducing 
differential cross sections  $d\sigma/d\Omega$ and vector analyzing power $A_y$ for $^4$He scattering 
in $E_{\rm in}=30 \sim 200$~MeV per nucleon~\cite{Toyokawa:2017pdd}. 
The success is true for proton scattering at $E_{\rm in}=65$~MeV~\cite{Toyokawa:2014yma}. 
We tested the Kyushu $g$-matrix folding model for $^{12}$C scattering on $^{9}$Be,  $^{12}$C, $^{27}$Al targets 
and found that the model is reliable 
in $30  \lsim E_{\rm in} \lsim 100 $~MeV and $250  \lsim E_{\rm in} \lsim 400 $~MeV~\cite{Tagami:2019svt}.

In this paper, therefore, we use the Kyushu $g$-matrix folding model not only for p+$^{48}$Ca scattering 
in $E_{\rm in}=30 \sim 48$~MeV but also for $^{48}$Ca+$^{12}$C scattering at $E_{\rm in}=280$~MeV 
per nucleon. 
The proton and neutron densities, $\rho_p(r)$ and $\rho_n(r)$, 
are scaled from the D1M-GHFB+AMP densities. As a way of taking the center-of-mass correction to the D1M-GHFB+AMP densities,  we use the method of Ref.~\cite{PRC.85.064613}, 
since the procedure is quite simple. 

\subsubsection{Single folding model for p+$^{48}$Ca scattering}
\label{Sec:Single folding model}

The potential $U(\vR)$ is composed of the direct and exchange parts,
$U^{\rm DR}(\vR)$ and $U^{\rm EX}(\vR)$~\cite{Minomo:2009ds,Watanabe:2014zea}: Namely, 
\bea
U(\vR)=U^{\rm DR}(\vR)+U^{\rm EX}(\vR) 
\label{eq:NA-potential}
\eea
with 
\bea
\label{eq:UD}
U^{\rm DR}(\vR) \hspace*{-0.15cm} &=& \hspace*{-0.15cm}
\sum_{\nu}\int             \rho^{\nu}_{\rm T}(\vrr_{\rm T})
            g^{\rm DR}_{\mu\nu}(s;\rho_{\mu\nu}) d \vrr_{\rm T}, \\
\label{eq:UEX}
U^{\rm EX}(\vR) \hspace*{-0.15cm} &=& \hspace*{-0.15cm}\sum_{\nu}
\int \rho^{\nu}_{\rm T}(\vrr_{\rm T},\vrr_{\rm T}+\vs) \nonumber \\
            &&~~\hspace*{-0.5cm}\times g^{\rm EX}_{\mu\nu}(s;\rho_{\mu\nu}) \exp{[-i\vK(\vR) \cdot \vs/M]}
             d \vrr_{\rm T},~~~~
            \label{U-EX-single}
\eea
where $\mu=-1/2$, the coordinate $\vR$ stands for the relative coordinate between an incident nucleon 
and a target (T), and 
$\vs \equiv -\vrr_{\rm T}+\vR$ for  the coordinate $\vrr_{\rm T}$ of the interacting nucleon from T.
Each of $\mu$ and $\nu$ denotes the $z$-component of isospin; $1/2$ means neutron and $-1/2$ does proton.
The nonlocal $U^{\rm EX}$ has been localized in Eq.~\eqref{U-EX-single}
with the local semi-classical approximation~\cite{Brieva-Rook-1,Brieva-Rook-2,Brieva-Rook-3},
where \vK(\vR) is the local momentum between the incident proton and T, and
$M= A_{\rm T}/(1 +A_{\rm T})$ for the target mass number $A_{\rm T}$. 
The validity of this localization is shown in Ref.~\cite{Minomo:2009ds}.

The direct and exchange parts, $g^{\rm DR}_{\mu\nu}$ and
$g^{\rm EX}_{\mu\nu}$, of the $g$-matrix depend on the local density
\bea
 \rho_{\mu\nu}=\sigma^{\mu} \rho^{\nu}_{\rm T}(\vrr_{\rm T}+\vs/2)
\label{local-density approximation}
\eea
at the midpoint of the interacting nucleon pair, where $\sigma^{\mu}$ is the Pauli matrix of an 
incident proton.

The direct and exchange parts, $g^{\rm DR}_{\mu\nu}$ and 
$g^{\rm EX}_{\mu\nu}$, of the $g$-matrix  are described by
\begin{align}
&\hspace*{0.5cm} g_{\mu\nu}^{\rm DR}(s;\rho_{\mu\nu}) \nonumber \\ 
&=
\begin{cases}
\displaystyle{\frac{1}{4} \sum_S} \hat{S}^2 g_{\mu\nu}^{S1}
 (s;\rho_{\mu\nu}) \hspace*{0.42cm} ; \hspace*{0.2cm} 
 {\rm for} \hspace*{0.1cm} \mu+\nu = \pm 1 
 \vspace*{0.2cm}\\
\displaystyle{\frac{1}{8} \sum_{S,T}} 
\hat{S}^2 g_{\mu\nu}^{ST}(s;\rho_{\mu\nu}), 
\hspace*{0.2cm} ; \hspace*{0.2cm} 
{\rm for} \hspace*{0.1cm} \mu+\nu = 0 
\end{cases}
\\
&\hspace*{0.5cm}
g_{\mu\nu}^{\rm EX}(s;\rho_{\mu\nu}) \nonumber \\
&=
\begin{cases}
\displaystyle{\frac{1}{4} \sum_S} (-1)^{S+1} 
\hat{S}^2 g_{\mu\nu}^{S1} (s;\rho_{\mu\nu}) 
\hspace*{0.34cm} ; \hspace*{0.2cm} 
{\rm for} \hspace*{0.1cm} \mu+\nu = \pm 1 \vspace*{0.2cm}\\
\displaystyle{\frac{1}{8} \sum_{S,T}} (-1)^{S+T} 
\hat{S}^2 g_{\mu\nu}^{ST}(s;\rho_{\mu\nu}) 
\hspace*{0.2cm} ; \hspace*{0.2cm}
{\rm for} \hspace*{0.1cm} \mu+\nu = 0 ~~~~~
\end{cases}
\end{align}
where $\hat{S} = {\sqrt {2S+1}}$ and $g_{\mu\nu}^{ST}$ are 
the spin-isospin components of the $g$-matrix.

\subsubsection{Double folding model for $^{48}$Ca+$^{12}$C scattering}
\label{Sec:Double folding model}

For nucleus-nucleus (P+T) scattering,  the potential $U(\vR)$ 
is $U^{\rm DR}+U^{\rm EX}$ defined by
\bea
\label{eq:UD}
U^{\rm DR}(\vR) \hspace*{-0.15cm} &=& \hspace*{-0.15cm} 
\sum_{\mu,\nu}\int \rho^{\mu}_{\rm P}(\vrr_{\rm P}) 
            \rho^{\nu}_{\rm T}(\vrr_{\rm T})
            g^{\rm DR}_{\mu\nu}(s;\rho_{\mu\nu}) d \vrr_{\rm P} d \vrr_{\rm T}, \\
\label{eq:UEX}
U^{\rm EX}(\vR) \hspace*{-0.15cm} &=& \hspace*{-0.15cm}\sum_{\mu,\nu} 
\int \rho^{\mu}_{\rm P}(\vrr_{\rm P},\vrr_{\rm P}-\vs)
\rho^{\nu}_{\rm T}(\vrr_{\rm T},\vrr_{\rm T}+\vs) \nonumber \\
            &&~~\hspace*{-0.5cm}\times g^{\rm EX}_{\mu\nu}(s;\rho_{\mu\nu}) \exp{[-i\vK(\vR) \cdot \vs/M_A]}
            d \vrr_{\rm P} d \vrr_{\rm T},~~~~
            \label{U-EX}
\eea
where $\vs=\vrr_{\rm P}-\vrr_{\rm T}+\vR$ 
for the coordinate $\vR$ between P and T. The coordinate 
$\vrr_{\rm P}$ 
($\vrr_{\rm T}$) denotes the location of an interacting nucleon 
from the center-of-mass of the projectile (target). 
The original form of $U^{\rm EX}$ is a non-local function of $\vR$,
but  it has been localized in Eq.~\eqref{U-EX}
with the local semi-classical approximation~\cite{Brieva-Rook-1,Brieva-Rook-2,Brieva-Rook-3} where 
P is assumed to propagate as a plane wave with
the local momentum $\hbar \vK(\vR)$ within a short range of the 
nucleon-nucleon interaction, where $M_A=A A_{\rm T}/(A +A_{\rm T})$
for the mass number $A$ ($A_{\rm T}$) of P (T).

The direct and exchange parts, $g^{\rm DR}_{\mu\nu}$ and 
$g^{\rm EX}_{\mu\nu}$, of the effective nucleon-nucleon interaction 
($g$-matrix) are assumed to depend on the local density
\bea
 \rho_{\mu\nu}=\rho^{\mu}_{\rm P}(\vrr_{\rm P}-\vs/2)
 +\rho^{\nu}_{\rm T}(\vrr_{\rm T}+\vs/2)
\label{local-density approximation}
\eea
at the midpoint of the interacting nucleon pair.
As for $^{12}$C, we use a phenomenological density of Ref.~\cite{C12-density}.

The relation $\sigma_{\rm R} \approx \sigma_{\rm I}+18.5$~mb  is shown 
in Ref.~\cite{Takechi:2020snn}. 
The difference hardly affects the resulting skin values.

\subsection{Scaling of proton and neutron densities}
\label{Sec:Scaling of proton and neutron densities}

We consider proton and neutron densities calculated with of  D1S-GHFB+AMP and D1M-GHFB+AMP as the original density $\rho(\vrr)$. 
The scaled density $\rho_{\rm scaling}(\vrr)$ is determined  
from the original density $\rho(\vrr)$ as
\bea
\rho_{\rm scaling}(\vrr) \equiv \frac{1}{\a^3}\rho(\vrr/\a), ~~\vrr_{\rm scaling} \equiv \vrr/\a
\label{eq:scaling}
\eea
with a scaling factor
\bea
\a=\sqrt{ \frac{\langle \vrr^2 \rangle_{\rm scaling}}{\langle \vrr^2 \rangle}} .
\eea
In Eq.~\eqref {eq:scaling}, we have replaced $\vrr$ by $\vrr/\a$ in the original density. 
Eventually, $\vrr$ dependence 
of   $\rho_{\rm scaling}(\vrr)$ is different from that of  $\rho(\vrr)$. 
We have multiplied the original density by $\a^{-3}$ 
 in order to normalize the scaled density. 
The symbol means $\sqrt{\langle \vrr^2 \rangle_{\rm scaling}}$ is the root-mean-square
radius of  $\rho_{\rm scaling}(\vrr)$. 

For later convenience, we refer to the proton (neutron) radius of the scaled proton (neutron) density $\rho^{\rm p}_{\rm scaling}(\vrr)$ 
($\rho^{\rm n}_{\rm scaling}(\vrr)$) as $r_{\rm p}({\rm scaling})$ ($r_{\rm n}({\rm scaling})$). 

When we scale the original D1S proton and neutron densities to the scaled densities base on 
 $r_{\rm skin}^{48}(E1{\rm pE}) =0.14 \sim 0.20~{\rm fm}$, 
 the value of $\a$ is 0.9992 for neutron and  0.9899 for proton. 
These values are quite close to 1.

\subsection{Structure models}
\label{Sec:Theoretical framework}
 
We recapitulate GHFB and GFHB+AMP, 
and explain GCM ($\beta$ mixing) and particle number variation after projection (PNVAP).

\subsubsection{GCM and PNVAP}
\label{Sec:GCM and PNVAP}

In PNVAP, the total energy is defined by 
\begin{align}
 E_{NZ}&\equiv \frac{\langle \Phi | \hat H \hat P_Z \hat P_N |\Phi\rangle}{\langle \Phi | \hat P_N \hat P_Z |\Phi\rangle}
\end{align}
with 
\begin{align}
 \delta E_{NZ}&=0,
\end{align}
where $P_N$ ($P_Z$) is the number projector for neutron (proton) and the intrinsic ground-state 
$|\Phi\rangle$ 
satisfies  ${\hat a_k}|\Phi\rangle=0$ for the annihilation operator ${\hat a_k}$ of quasi-particle for any $k$.

In GCM, we consider the following Hamiltonian and norm matrices as 
\begin{align}
 \left\{\begin{matrix}
  {\cal H}^I_{Kn,K'n'} \cr {\cal N}^I_{Kn,K'n'}
 \end{matrix}\right\}&\equiv \langle \Phi_n |
  \left\{\begin{matrix}
  \hat H \cr 1
 \end{matrix}\right\} \hat P^I_{KK'}  |\Phi_{n'}\rangle
 \\
 {\rm with}~~~~~~~~~~~~~~~~~~~~~~~~~~~~~~& \notag \\
 |\Phi_n \rangle= |\Phi(q_{20,n})\rangle&
 ,  
 \langle \Phi(q_{20,n})| \hat Q_{20}  |\Phi(q_{20,n})\rangle=q_{20,n};
\end{align}
in actual calculations, the $\{q_{20,n}\}$ are taken from -60~fm$^2$ to 140~fm$^2$ at an interval of 20~fm$^2$. 
The total energy with the total spin $I$ is 
\begin{align}
 E_I=\frac{\sum_{KnK'n'}g^*_{IKn} {\cal H}^I_{Kn,K'n'} g_{IK'n'}}{\sum_{KnK'n'}g^*_{IKn} {\cal N}^I_{Kn,K'n'} g_{IK'n'}}&
 \\
{\rm under~the~conditions}~~~~~~~~~~~~~~~~~~~~~~~~~~~~~~~~~~~~~~~~& \notag \\
 \frac{\delta E_I}{\delta g^*_{IKn}}&=0,
 \\
 \sum_{K'n'}\left[ {\cal H}^I_{Kn,K'n'} g_{IK'n'}-E_I {\cal N}^I_{Kn,K'n'}\right] g_{IK'n'}&=0.
\end{align}

\subsubsection{GFHB+AMP}
\label{Sec:GFHB+AMP}

In GHFB+AMP, the total wave function  $| \Psi^I_{M} \rangle$ with the AMP is defined by 
\begin{equation}
 | \Psi^I_{M} \rangle =
 \sum_{K, n=0}^{N} g^I_{K n} \hat P^I_{MK}|\Phi_n \rangle ,
\label{eq:prjc}
\end{equation}
where $\hat P^I_{MK}$ is the angular-momentum-projector and the 
$|\Phi_n \rangle$ for $n=0,1,\cdots,N$ are mean-field (GHFB) states, 
where $N+1$ is the number  of the  states.  
The coefficients $g^I_{K n}$ are obtained by solving the Hill-Wheeler equation
\begin{equation}
 \sum_{K^\prime n^\prime }{\cal H}^I_{Kn,K^\prime n^\prime }\ g^I_{K^\prime n^\prime } =
 E_I\,
 \sum_{K^\prime n^\prime }{\cal N}^I_{Kn,K^\prime n^\prime }\ g^I_{K^\prime n^\prime },
\end{equation}
with the Hamiltonian and norm kernels defined by
\begin{equation}
 \left\{ \begin{array}{c}
   {\cal H}^I_{Kn,K^\prime n^\prime } \\ {\cal N}^I_{Kn,K^\prime n^\prime } \end{array}
 \right\} = \langle \Phi_n |
 \left\{ \begin{array}{c}
   \hat{H} \\ 1 \end{array}
 \right\} \hat{P}_{KK^\prime }^I | \Phi_{n'} \rangle.
\end{equation}

For even nuclei, there is no blocking state, i.e., $N=0$ in the Hill-Wheeler equation. 
We can thus perform  GHFB+AMP. 
However,  we have to find the value of $\beta$ at 
which the ground-state energy becomes minimum.  
In this step, the AMP has to be performed for any $\beta$, so that the Hill-Wheeler calculation is  heavy. 
In fact, the AMP is not taken for mean field calculations in many works. 
The reason why we do not take into account $\gamma$ deformation is 
that the deformation does not affect  $\sigma_{\rm R}$~\cite{Sumi:2012fr}.

\begin{figure}[htb]
\centering
\vspace{0cm}
\includegraphics[width=0.4\textwidth]{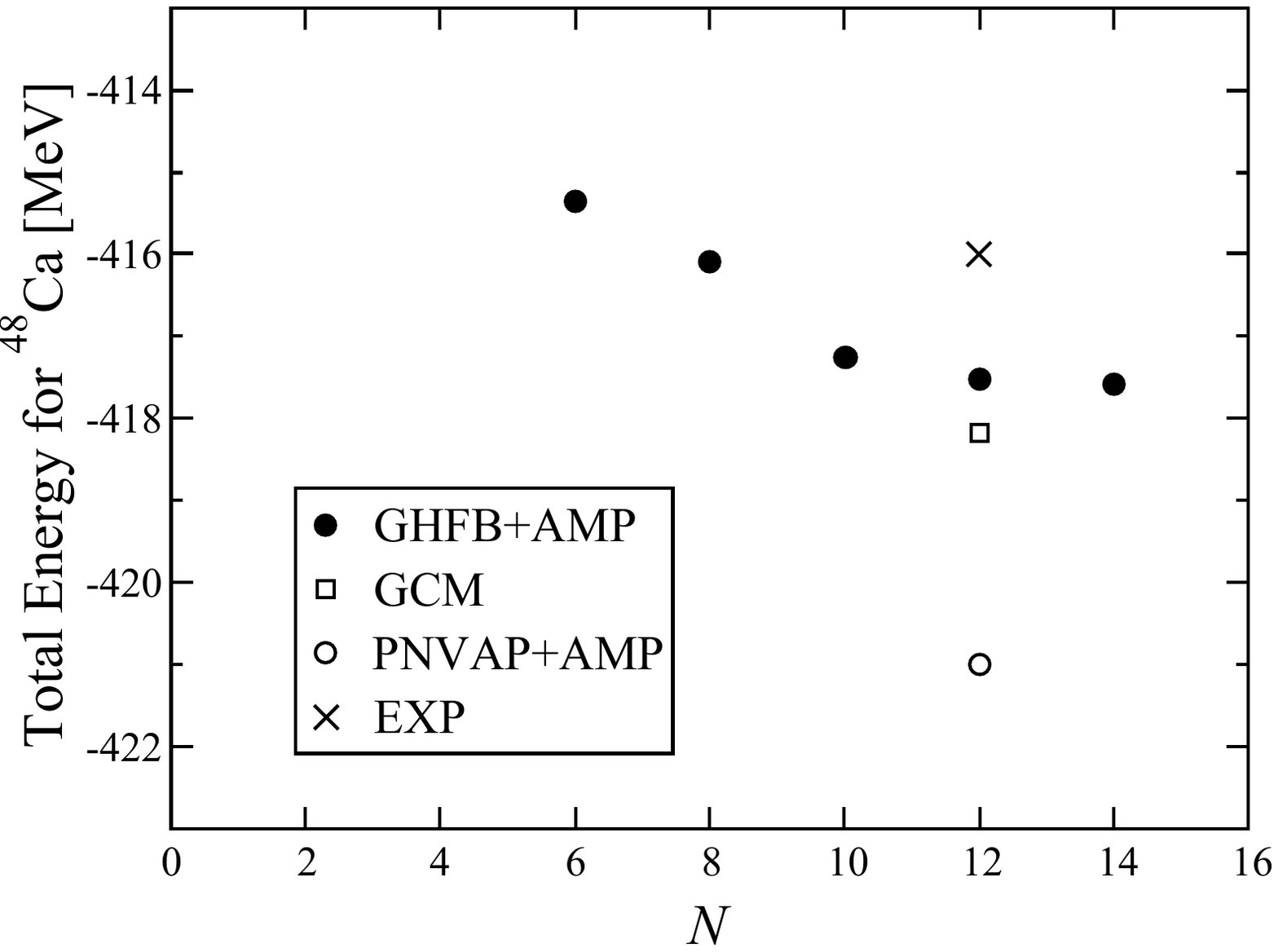}
\includegraphics[width=0.4\textwidth]{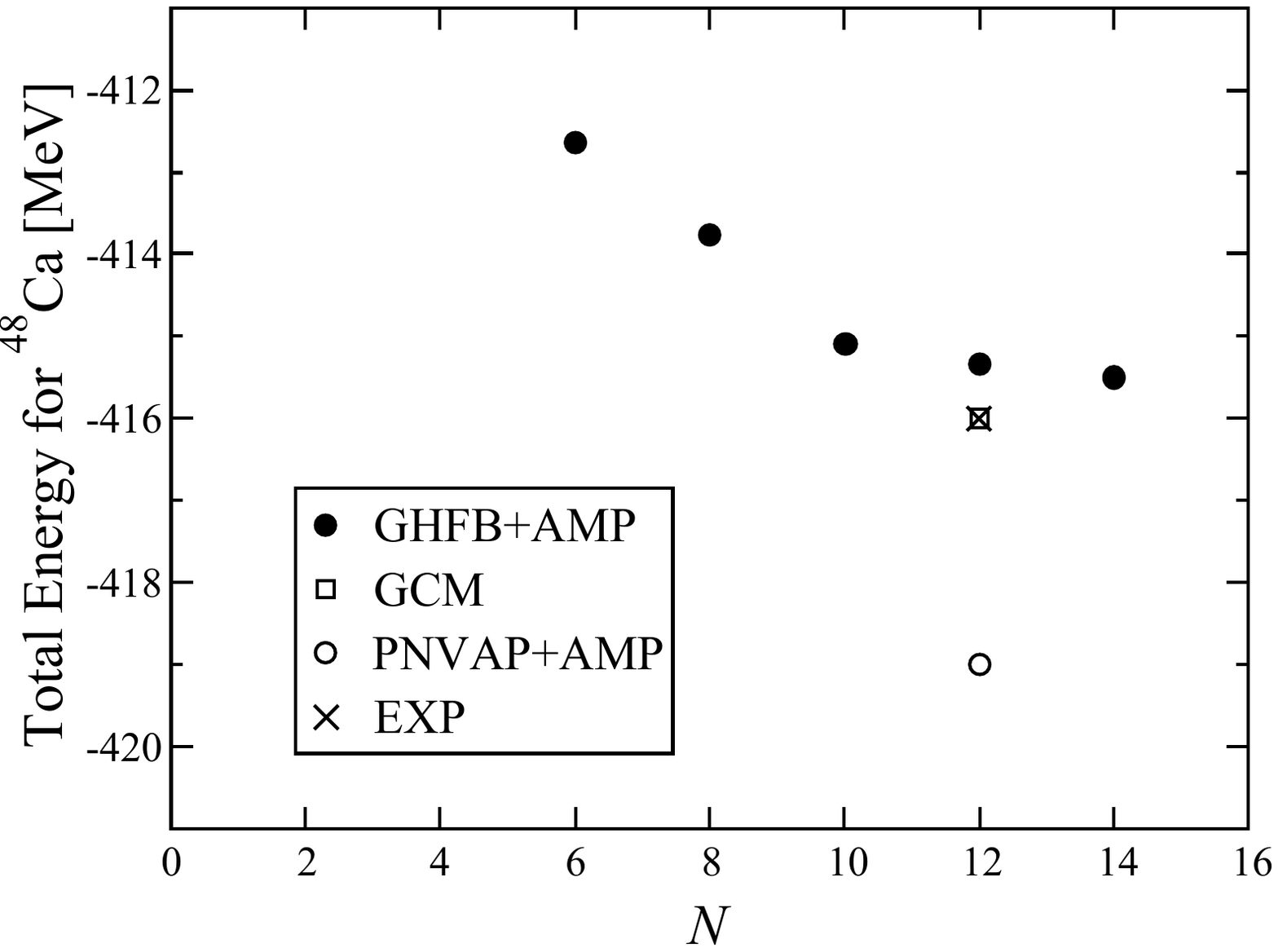}
\vspace{-10pt}
\caption{$N$ dependence of  total energy for $^{48}$Ca; the upper (lower) panel shows the results for D1S (D1M). 
Closed  circles denote the result of GHFB+AMP, while  the open circle  corresponds 
to the results of GHFB+AMP+PNVAP.  A square stands for the result of GHFB+AMP+GCM.
In the lower panel, the data coincidences with the result of GHFB+AMP+GCM.
Experimental data (cross) is taken from Ref.~\cite{HP:NuDat2.8}. 
}
\label{fig:N dependence of TE}
\end{figure}

Figure \ref{fig:N dependence of TE} shows $N$ dependence 
of total energy for $^{48}$Ca, 
where we take  D1S (upper panel) and D1M (lower panel). 
The calculated total energy saturates at $N=12$. 
We then take into account effects of GCM and  PNVAP in GHFB+AMP for $N=12$.  
Effects of GCM ($\beta$ mixing) and PNVAP are small. 
The D1M-GHFB+AMP result is closer to the data~\cite{HP:NuDat2.8} than 
the D1S-GHFB+AMP result. This is one of the reason why we use D1M-GHFB+AMP as the densities for $^{48}$Ca.

\section{Results}
\label{Sec:Results}

\subsection{Reactions}

\begin{figure}[H]
\centering
\vspace{0cm}
\includegraphics[width=0.45\textwidth]{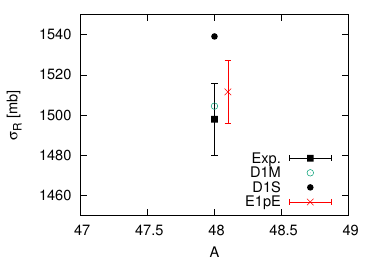}
\caption{
Reaction cross sections for  $^{48}$Ca+$^{12}$C scattering in $E_{\rm in}=280$~MeV 
per nucleon. 
Close circle denotes the result of D1S-GHFB+AMP. 
The result based on $r_{\rm skin}^{48}(E1{\rm pE})$ is shown at $A=48.1$ instead of  $A=48$. 
The data are taken from Ref.~\cite{Tanaka:2019pdo} on $\sigma_{\rm I}$. 
}
\label{fig:48Ca+12C.pdf}
\end{figure}

As shown in Fig~\ref{fig:48Ca+12C.pdf} for  $^{48}$Ca+$^{12}$C scattering at $E_{\rm in}=280$~MeV 
per nucleon,  the result of D1M-GHFB+AMP 
reproduces the data~\cite{Tanaka:2019pdo} on $\sigma_{\rm I}$. 
The result of D1S-GHFB+AMP overshoots the experimental data.~\cite{Tanaka:2019pdo}.

\begin{figure}[H]
\centering
\vspace{0cm}
\includegraphics[width=0.45\textwidth]{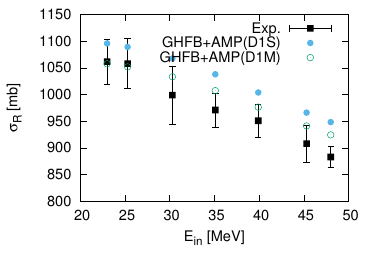}
\caption{
Reaction cross sections for  p+$^{48}$Ca scattering in $E_{\rm in}=23 \sim 48$~MeV. 
Open circles stand for the results of D1M-GHFB+AMP, whereas closed circles  correspond to 
those of   D1S-GHFB+AMP. 
The data are taken from Ref.~\cite{Carlson:1994fq}. 
}
\label{fig:p+48Ca-0.pdf}

\end{figure}

Figure \ref{fig:p+48Ca-0.pdf} shows reaction cross sections $\sigma_{\rm R}$ 
for p+$^{48}$Ca scattering in $E_{\rm in}=23 \sim 48$~MeV. 
In $E_{\rm in}=30 \sim 48$~MeV where the Kyushu $g$-matrix model is reliable, 
the results of D1M-GHFB+AMP yield better agreement with the data~\cite{Carlson:1994fq} 
than those of  D1S-GHFB+AMP. 
Now we consider  D1M-GHFB+AMP.

We first deduce neutron radius $r_{\rm n}(E1{\rm pE})=3.525 \sim 3.585~{\rm fm}$  
and matter radius  $r_{\rm m}(E1{\rm pE})=3.467 \sim 3.503~{\rm fm}$
from$r_{\rm skin}^{48}(E1{\rm pE}) =0.14 \sim 0.20~{\rm fm}$ and 
$r_p({\rm exp})=3.397$~fm~\cite{ADNDT.99.69} of electron scattering. 
Using Eq.~\eqref{eq:scaling}, we then scale the D1M-GHFB+AMP proton and neutron 
densities so as to  $r_{\rm n}(E1{\rm pE})=r_{\rm n}({\rm scaling})$ 
and $r_{\rm p}({\rm exp})=r_{\rm p}({\rm scaling})$. 
The result based on $r_{\rm skin}^{48}(E1{\rm pE})$ is shown at $A=48.1$ instead of  $A=48$ 
in Fig~\ref{fig:48Ca+12C.pdf}.

The result based on  $r_{\rm skin}^{48}(E1{\rm pE})$ 
is consistent with the data~\cite{Tanaka:2019pdo} on $\sigma_{\rm I}$, as shown 
in Fig~\ref{fig:48Ca+12C.pdf}.
We then scale the D1M-GHFB+AMP densities so 
that the  $\sigma_{\rm R}$ calculated from the scaled densities can agree with $\sigma_{\rm I}$ 
under the condition that $r_{p}({\rm scaling})=r_p({\rm exp})=3.397$~fm. 
The $r_{\rm m}({\rm exp})$ thus obtained is $3.456 \sim 3.526$~fm. 
We obtain $r_{\rm skin}^{48}({\rm exp})=0.101 \sim 0.219~{\rm fm}$ and 
$r_{n}({\rm exp})= 3.497 \sim 3.615~{\rm fm}$ from  the  $r_{\rm m}({\rm exp})$ and the 
$r_{p}({\rm exp})$.

\begin{figure}[H]
\centering
\vspace{0cm}
\includegraphics[width=0.45\textwidth]{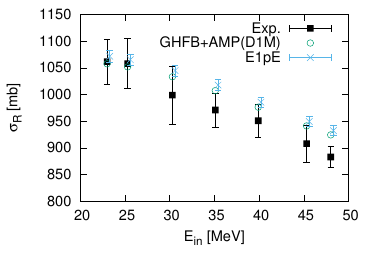}
\caption{
Reaction cross sections for  p+$^{48}$Ca scattering in $E_{\rm in}=23 \sim 48$~MeV. 
Open circles stand for the results of D1M-GHFB+AMP. 
The results based on $r_{\rm skin}^{48}(E1{\rm pE})$ are shown at $E_{\rm in}+0.3$ instead of  
$E_{\rm in}$. The data are taken from Ref.~\cite{Carlson:1994fq}. 
}
\label{fig:p+48Ca.pdf}
\end{figure}

Figure \ref{fig:p+48Ca.pdf} shows reaction cross sections $\sigma_{\rm R}$ 
for p+$^{48}$Ca scattering in $E_{\rm in}=23 \sim 48$~MeV. 
In $E_{\rm in}=30 \sim 48$~MeV where the Kyushu $g$-matrix model is reliable, 
the results of D1M-GHFB+AMP are near the upper bound of 
the data~\cite{Carlson:1994fq}, but the results  based on 
$r_{\rm skin}^{48}(E1{\rm pE})$ 
somewhat overshoot the upper bound of the data~\cite{Carlson:1994fq} on  $\sigma_{\rm R}$.

Since the results  based on $r_{\rm skin}^{48}(E1{\rm pE})$ 
are near the upper bound of the data~\cite{Carlson:1994fq}, 
we then scale the D1M-DHFB+AMP proton and neutron densities   
so that results of the scaled densities can reproduce the upper limit of $\sigma_{\rm R}({\rm exp})$ 
in $E_{\rm in}=30 \sim 48$~MeV 
under the condition that $r_{p}({\rm scaling})=r_p({\rm exp})=3.397$~fm.  
The $r_{\rm m}({\rm exp})$ thus obtained depend on $E_{\rm in}$. We then take 
the weighted mean and its error for five $E_{\rm in}$ in  $E_{\rm in}=30 \sim 48$~MeV. 
The resulting  $r_{\rm m}({\rm exp})$ is $3.490 \pm 0.025$~fm.  
We obtain $r_{\rm skin}^{48}({\rm exp})=0.158 \pm 0.025~{\rm fm}$ and 
$r_{n}({\rm exp})= 3.555 \pm 0.025~{\rm fm}$ from  the  $r_{\rm m}({\rm exp})$ and the $r_{p}({\rm exp})$.

Eventually, for D1M, we determine two values on $r_{\rm skin}^{48}({\rm exp})$;    
\bea
r_{\rm skin}^{48}({\rm exp})=0.158 \pm 0.025~{\rm fm}
\eea
for p+$^{48}$Ca scattering in  $E_{\rm in}=30 \sim 48$~MeV
and  
\bea
r_{\rm skin}^{48}({\rm exp})=0.160 	\pm 0.059~{\rm fm}
\eea
for  $^{48}$Ca + $^{12}$C scattering at 280~MeV per nucleon. 
Finally, we take the weighted mean and its error for two skin values mentioned above. 
Our skin value is 
\bea
r_{\rm skin}^{48}({\rm exp})=0.158 \pm	0.023~{\rm fm}
\label{final-48-skin-D1M}
\eea
for D1M.

For comparison, we consider D1S-GHFB+AMP. The same procedure is taken for D1S. The result  
\bea
r_{\rm skin}^{48}({\rm exp})=0.125 \pm	0.02~{\rm fm}. 
\label{final-48-skin-D1S}
\eea

For $^{48}$Ca, D1M is better than D1S.  
We then take the result of D1M+GHFB+AMP. 
Our final result is 
\bea
r_{\rm skin}^{48}({\rm exp})=0.158 \pm	0.023 \pm 0.012~{\rm fm},  
\label{final-48-skin-D1M-th}
\eea
where the first error is an experimental one and the second error is a theoretical error from D1M and D1S; see Table I for the skin values $r_{\rm skin}^{48}$ of D1M and D1S.

\subsection{Relation between $r_{\rm skin}^{48}$ and $L$}
\label{the lower bound}

Finally, we obtain  $L$ for neutron matter from $r_{\rm skin}^{48}$. For this purpose, we derive 
the  $r_{\rm skin}^{48}$-$L$ relation from the 206 EoSs of Table I. The relation is 
\bea
r_{\rm skin}^{48}=0.0009 L + 0.125 > 0.125~{\rm fm}
\label{final-48-skin-L}
\eea
with $R=0.98$, because of $L >0$. Equation \eqref{final-48-skin-L} indicates that 
the lower limit of $r_{\rm skin}^{48}$ is 0.125~fm. 

Using Eqs.~\eqref{final-48-skin-D1M-th} and \eqref{final-48-skin-L}, we finally obtain
\bea
L=0 \sim 76~~{\rm MeV}.  
\eea
The $L=0 \sim 76~{\rm MeV}$~MeV  deduced from 
$r_{\rm skin}^{48}({\rm exp})=0.158 \pm	0.023 \pm 0.012~{\rm fm}$ is smaller 
than   $L=76-165~{\rm MeV}$ from PREX2.  This is an interesting issue. 
The result of  CREX  will answer this issue.

\section*{Acknowledgements}
We thank Prof. Matsuzaki  and Prof. Yasutake for their comments.

\bibliography{Folding-v9}

\begin{thebibliography}{69}%
\makeatletter
\providecommand \@ifxundefined [1]{%
 \@ifx{#1\undefined}
}%
\providecommand \@ifnum [1]{%
 \ifnum #1\expandafter \@firstoftwo
 \else \expandafter \@secondoftwo
 \fi
}%
\providecommand \@ifx [1]{%
 \ifx #1\expandafter \@firstoftwo
 \else \expandafter \@secondoftwo
 \fi
}%
\providecommand \natexlab [1]{#1}%
\providecommand \enquote  [1]{``#1''}%
\providecommand \bibnamefont  [1]{#1}%
\providecommand \bibfnamefont [1]{#1}%
\providecommand \citenamefont [1]{#1}%
\providecommand \href@noop [0]{\@secondoftwo}%
\providecommand \href [0]{\begingroup \@sanitize@url \@href}%
\providecommand \@href[1]{\@@startlink{#1}\@@href}%
\providecommand \@@href[1]{\endgroup#1\@@endlink}%
\providecommand \@sanitize@url [0]{\catcode `\\12\catcode `\$12\catcode
  `\&12\catcode `\#12\catcode `\^12\catcode `\_12\catcode `\%12\relax}%
\providecommand \@@startlink[1]{}%
\providecommand \@@endlink[0]{}%
\providecommand \url  [0]{\begingroup\@sanitize@url \@url }%
\providecommand \@url [1]{\endgroup\@href {#1}{\urlprefix }}%
\providecommand \urlprefix  [0]{URL }%
\providecommand \Eprint [0]{\href }%
\providecommand \doibase [0]{http://dx.doi.org/}%
\providecommand \selectlanguage [0]{\@gobble}%
\providecommand \bibinfo  [0]{\@secondoftwo}%
\providecommand \bibfield  [0]{\@secondoftwo}%
\providecommand \translation [1]{[#1]}%
\providecommand \BibitemOpen [0]{}%
\providecommand \bibitemStop [0]{}%
\providecommand \bibitemNoStop [0]{.\EOS\space}%
\providecommand \EOS [0]{\spacefactor3000\relax}%
\providecommand \BibitemShut  [1]{\csname bibitem#1\endcsname}%
\let\auto@bib@innerbib\@empty
\bibitem [{\citenamefont {Horowitz}\ \emph {et~al.}(2001)\citenamefont
  {Horowitz}, \citenamefont {Pollock}, \citenamefont {Souder},\ and\
  \citenamefont {Michaels}}]{PRC.63.025501}%
  \BibitemOpen
  \bibfield  {author} {\bibinfo {author} {\bibfnamefont {C.~J.}\ \bibnamefont
  {Horowitz}}, \bibinfo {author} {\bibfnamefont {S.~J.}\ \bibnamefont
  {Pollock}}, \bibinfo {author} {\bibfnamefont {P.~A.}\ \bibnamefont {Souder}},
  \ and\ \bibinfo {author} {\bibfnamefont {R.}~\bibnamefont {Michaels}},\
  }\href {\doibase 10.1103/PhysRevC.63.025501} {\bibfield  {journal} {\bibinfo
  {journal} {Phys. Rev. C}\ }\textbf {\bibinfo {volume} {63}},\ \bibinfo
  {pages} {025501} (\bibinfo {year} {2001})}\BibitemShut {NoStop}%
\bibitem [{\citenamefont {Abrahamyan}\ \emph {et~al.}(2012)\citenamefont
  {Abrahamyan}, \citenamefont {Ahmed}, \citenamefont {Albataineh},
  \citenamefont {Aniol}, \citenamefont {Armstrong}, \citenamefont {Armstrong},
  \citenamefont {Averett}, \citenamefont {Babineau}, \citenamefont {Barbieri},
  \citenamefont {Bellini} \emph {et~al.}}]{PRL.108.112502}%
  \BibitemOpen
  \bibfield  {author} {\bibinfo {author} {\bibfnamefont {S.}~\bibnamefont
  {Abrahamyan}}, \bibinfo {author} {\bibfnamefont {Z.}~\bibnamefont {Ahmed}},
  \bibinfo {author} {\bibfnamefont {H.}~\bibnamefont {Albataineh}}, \bibinfo
  {author} {\bibfnamefont {K.}~\bibnamefont {Aniol}}, \bibinfo {author}
  {\bibfnamefont {D.~S.}\ \bibnamefont {Armstrong}}, \bibinfo {author}
  {\bibfnamefont {W.}~\bibnamefont {Armstrong}}, \bibinfo {author}
  {\bibfnamefont {T.}~\bibnamefont {Averett}}, \bibinfo {author} {\bibfnamefont
  {B.}~\bibnamefont {Babineau}}, \bibinfo {author} {\bibfnamefont
  {A.}~\bibnamefont {Barbieri}}, \bibinfo {author} {\bibfnamefont
  {V.}~\bibnamefont {Bellini}},  \emph {et~al.} (\bibinfo {collaboration} {PREX
  Collaboration}),\ }\href {\doibase 10.1103/PhysRevLett.108.112502} {\bibfield
   {journal} {\bibinfo  {journal} {Phys. Rev. Lett.}\ }\textbf {\bibinfo
  {volume} {108}},\ \bibinfo {pages} {112502} (\bibinfo {year}
  {2012})}\BibitemShut {NoStop}%
\bibitem [{\citenamefont {Horowitz}\ \emph {et~al.}(2012)\citenamefont
  {Horowitz}, \citenamefont {Ahmed}, \citenamefont {Jen}, \citenamefont
  {Rakhman}, \citenamefont {Souder}, \citenamefont {Dalton}, \citenamefont
  {Liyanage}, \citenamefont {Paschke}, \citenamefont {Saenboonruang},
  \citenamefont {Silwal}, \citenamefont {Franklin}, \citenamefont {Friend},
  \citenamefont {Quinn}, \citenamefont {Kumar}, \citenamefont {McNulty},
  \citenamefont {Mercado}, \citenamefont {Riordan}, \citenamefont {Wexler},
  \citenamefont {Michaels},\ and\ \citenamefont {Urciuoli}}]{PRC.85.032501}%
  \BibitemOpen
  \bibfield  {author} {\bibinfo {author} {\bibfnamefont {C.~J.}\ \bibnamefont
  {Horowitz}}, \bibinfo {author} {\bibfnamefont {Z.}~\bibnamefont {Ahmed}},
  \bibinfo {author} {\bibfnamefont {C.-M.}\ \bibnamefont {Jen}}, \bibinfo
  {author} {\bibfnamefont {A.}~\bibnamefont {Rakhman}}, \bibinfo {author}
  {\bibfnamefont {P.~A.}\ \bibnamefont {Souder}}, \bibinfo {author}
  {\bibfnamefont {M.~M.}\ \bibnamefont {Dalton}}, \bibinfo {author}
  {\bibfnamefont {N.}~\bibnamefont {Liyanage}}, \bibinfo {author}
  {\bibfnamefont {K.~D.}\ \bibnamefont {Paschke}}, \bibinfo {author}
  {\bibfnamefont {K.}~\bibnamefont {Saenboonruang}}, \bibinfo {author}
  {\bibfnamefont {R.}~\bibnamefont {Silwal}}, \bibinfo {author} {\bibfnamefont
  {G.~B.}\ \bibnamefont {Franklin}}, \bibinfo {author} {\bibfnamefont
  {M.}~\bibnamefont {Friend}}, \bibinfo {author} {\bibfnamefont
  {B.}~\bibnamefont {Quinn}}, \bibinfo {author} {\bibfnamefont {K.~S.}\
  \bibnamefont {Kumar}}, \bibinfo {author} {\bibfnamefont {D.}~\bibnamefont
  {McNulty}}, \bibinfo {author} {\bibfnamefont {L.}~\bibnamefont {Mercado}},
  \bibinfo {author} {\bibfnamefont {S.}~\bibnamefont {Riordan}}, \bibinfo
  {author} {\bibfnamefont {J.}~\bibnamefont {Wexler}}, \bibinfo {author}
  {\bibfnamefont {R.~W.}\ \bibnamefont {Michaels}}, \ and\ \bibinfo {author}
  {\bibfnamefont {G.~M.}\ \bibnamefont {Urciuoli}},\ }\href {\doibase
  10.1103/PhysRevC.85.032501} {\bibfield  {journal} {\bibinfo  {journal} {Phys.
  Rev. C}\ }\textbf {\bibinfo {volume} {85}},\ \bibinfo {pages} {032501}
  (\bibinfo {year} {2012})}\BibitemShut {NoStop}%
\bibitem [{\citenamefont {Adhikari}\ \emph {et~al.}(2021)\citenamefont
  {Adhikari} \emph {et~al.}}]{Adhikari:2021phr}%
  \BibitemOpen
  \bibfield  {author} {\bibinfo {author} {\bibfnamefont {D.}~\bibnamefont
  {Adhikari}} \emph {et~al.} (\bibinfo {collaboration} {PREX}),\ }\href
  {\doibase 10.1103/PhysRevLett.126.172502} {\bibfield  {journal} {\bibinfo
  {journal} {Phys. Rev. Lett.}\ }\textbf {\bibinfo {volume} {126}},\ \bibinfo
  {pages} {172502} (\bibinfo {year} {2021})},\ \Eprint
  {http://arxiv.org/abs/2102.10767} {arXiv:2102.10767 [nucl-ex]} \BibitemShut
  {NoStop}%
\bibitem [{\citenamefont {Michaels}\ \emph {et~al.}(2005)\citenamefont
  {Michaels} \emph {et~al.}}]{PREX05}%
  \BibitemOpen
  \bibfield  {author} {\bibinfo {author} {\bibfnamefont {R.}~\bibnamefont
  {Michaels}} \emph {et~al.},\ }\href@noop {} {\enquote {\bibinfo {title} {Lead
  radius experiment prex proposal},}\ }\bibinfo {howpublished}
  {http://hallaweb.jlab.org/parity/prex/} (\bibinfo {year} {2005})\BibitemShut
  {NoStop}%
\bibitem [{\citenamefont {Tamii}\ \emph
  {et~al.}(2011{\natexlab{a}})\citenamefont {Tamii} \emph
  {et~al.}}]{Tamii:2011pv}%
  \BibitemOpen
  \bibfield  {author} {\bibinfo {author} {\bibfnamefont {A.}~\bibnamefont
  {Tamii}} \emph {et~al.},\ }\href {\doibase 10.1103/PhysRevLett.107.062502}
  {\bibfield  {journal} {\bibinfo  {journal} {Phys. Rev. Lett.}\ }\textbf
  {\bibinfo {volume} {107}},\ \bibinfo {pages} {062502} (\bibinfo {year}
  {2011}{\natexlab{a}})},\ \Eprint {http://arxiv.org/abs/1104.5431}
  {arXiv:1104.5431 [nucl-ex]} \BibitemShut {NoStop}%
\bibitem [{\citenamefont {Birkhan}\ \emph {et~al.}(2017)\citenamefont {Birkhan}
  \emph {et~al.}}]{Birkhan:2016qkr}%
  \BibitemOpen
  \bibfield  {author} {\bibinfo {author} {\bibfnamefont {J.}~\bibnamefont
  {Birkhan}} \emph {et~al.},\ }\href {\doibase 10.1103/PhysRevLett.118.252501}
  {\bibfield  {journal} {\bibinfo  {journal} {Phys. Rev. Lett.}\ }\textbf
  {\bibinfo {volume} {118}},\ \bibinfo {pages} {252501} (\bibinfo {year}
  {2017})},\ \Eprint {http://arxiv.org/abs/1611.07072} {arXiv:1611.07072
  [nucl-ex]} \BibitemShut {NoStop}%
\bibitem [{\citenamefont {Carlson}\ \emph {et~al.}(1994)\citenamefont
  {Carlson}, \citenamefont {Cox}, \citenamefont {Davison}, \citenamefont
  {Eliyakut-Roshko}, \citenamefont {McCamis},\ and\ \citenamefont
  {Oers}}]{Carlson:1994fq}%
  \BibitemOpen
  \bibfield  {author} {\bibinfo {author} {\bibfnamefont {R.~F.}\ \bibnamefont
  {Carlson}}, \bibinfo {author} {\bibfnamefont {A.~J.}\ \bibnamefont {Cox}},
  \bibinfo {author} {\bibfnamefont {N.~E.}\ \bibnamefont {Davison}}, \bibinfo
  {author} {\bibfnamefont {T.}~\bibnamefont {Eliyakut-Roshko}}, \bibinfo
  {author} {\bibfnamefont {R.~H.}\ \bibnamefont {McCamis}}, \ and\ \bibinfo
  {author} {\bibfnamefont {W.~T. H.~v.}\ \bibnamefont {Oers}},\ }\href
  {\doibase 10.1103/PhysRevC.49.3090} {\bibfield  {journal} {\bibinfo
  {journal} {Phys. Rev. C}\ }\textbf {\bibinfo {volume} {49}},\ \bibinfo
  {pages} {3090} (\bibinfo {year} {1994})}\BibitemShut {NoStop}%
\bibitem [{\citenamefont {Tanaka}\ \emph {et~al.}(2020)\citenamefont {Tanaka}
  \emph {et~al.}}]{Tanaka:2019pdo}%
  \BibitemOpen
  \bibfield  {author} {\bibinfo {author} {\bibfnamefont {M.}~\bibnamefont
  {Tanaka}} \emph {et~al.},\ }\href {\doibase 10.1103/PhysRevLett.124.102501}
  {\bibfield  {journal} {\bibinfo  {journal} {Phys. Rev. Lett.}\ }\textbf
  {\bibinfo {volume} {124}},\ \bibinfo {pages} {102501} (\bibinfo {year}
  {2020})},\ \Eprint {http://arxiv.org/abs/1911.05262} {arXiv:1911.05262
  [nucl-ex]} \BibitemShut {NoStop}%
\bibitem [{\citenamefont {Novario}\ \emph {et~al.}(2020)\citenamefont
  {Novario}, \citenamefont {Hagen}, \citenamefont {Jansen},\ and\ \citenamefont
  {Papenbrock}}]{PRC.102.051303}%
  \BibitemOpen
  \bibfield  {author} {\bibinfo {author} {\bibfnamefont {S.~J.}\ \bibnamefont
  {Novario}}, \bibinfo {author} {\bibfnamefont {G.}~\bibnamefont {Hagen}},
  \bibinfo {author} {\bibfnamefont {G.~R.}\ \bibnamefont {Jansen}}, \ and\
  \bibinfo {author} {\bibfnamefont {T.}~\bibnamefont {Papenbrock}},\ }\href
  {\doibase 10.1103/PhysRevC.102.051303} {\bibfield  {journal} {\bibinfo
  {journal} {Phys. Rev. C}\ }\textbf {\bibinfo {volume} {102}},\ \bibinfo
  {pages} {051303} (\bibinfo {year} {2020})}\BibitemShut {NoStop}%
\bibitem [{\citenamefont {Shen}\ \emph {et~al.}(2020)\citenamefont {Shen},
  \citenamefont {Ji}, \citenamefont {Hu},\ and\ \citenamefont
  {Sumiyoshi}}]{AJ.891.148}%
  \BibitemOpen
  \bibfield  {author} {\bibinfo {author} {\bibfnamefont {H.}~\bibnamefont
  {Shen}}, \bibinfo {author} {\bibfnamefont {F.}~\bibnamefont {Ji}}, \bibinfo
  {author} {\bibfnamefont {J.}~\bibnamefont {Hu}}, \ and\ \bibinfo {author}
  {\bibfnamefont {K.}~\bibnamefont {Sumiyoshi}},\ }\href {\doibase
  10.3847/1538-4357/ab72fd} {\bibfield  {journal} {\bibinfo  {journal}
  {Astrophys. J.}\ }\textbf {\bibinfo {volume} {891}},\ \bibinfo {pages} {148}
  (\bibinfo {year} {2020})}\BibitemShut {NoStop}%
\bibitem [{\citenamefont {Horowitz}(2019)}]{AP.411.167992}%
  \BibitemOpen
  \bibfield  {author} {\bibinfo {author} {\bibfnamefont {C.}~\bibnamefont
  {Horowitz}},\ }\href {\doibase https://doi.org/10.1016/j.aop.2019.167992}
  {\bibfield  {journal} {\bibinfo  {journal} {Ann. Phys. (Amsterdam)}\ }\textbf
  {\bibinfo {volume} {411}},\ \bibinfo {pages} {167992} (\bibinfo {year}
  {2019})}\BibitemShut {NoStop}%
\bibitem [{\citenamefont {{Wei, Jin-Biao}}\ \emph {et~al.}(2020)\citenamefont
  {{Wei, Jin-Biao}}, \citenamefont {{Lu, Jia-Jing}}, \citenamefont {{Burgio, G.
  F.}}, \citenamefont {{Li, Zeng-Hua}},\ and\ \citenamefont {{Schulze,
  H.-J.}}}]{EPJA.56.63}%
  \BibitemOpen
  \bibfield  {author} {\bibinfo {author} {\bibnamefont {{Wei, Jin-Biao}}},
  \bibinfo {author} {\bibnamefont {{Lu, Jia-Jing}}}, \bibinfo {author}
  {\bibnamefont {{Burgio, G. F.}}}, \bibinfo {author} {\bibnamefont {{Li,
  Zeng-Hua}}}, \ and\ \bibinfo {author} {\bibnamefont {{Schulze, H.-J.}}},\
  }\href {\doibase 10.1140/epja/s10050-020-00058-3} {\bibfield  {journal}
  {\bibinfo  {journal} {Eur. Phys. J. A}\ }\textbf {\bibinfo {volume} {56}},\
  \bibinfo {pages} {63} (\bibinfo {year} {2020})}\BibitemShut {NoStop}%
\bibitem [{\citenamefont {Thiel}\ \emph {et~al.}(2019)\citenamefont {Thiel},
  \citenamefont {Sfienti}, \citenamefont {Piekarewicz}, \citenamefont
  {Horowitz},\ and\ \citenamefont {Vanderhaeghen}}]{JPG.46.093003}%
  \BibitemOpen
  \bibfield  {author} {\bibinfo {author} {\bibfnamefont {M.}~\bibnamefont
  {Thiel}}, \bibinfo {author} {\bibfnamefont {C.}~\bibnamefont {Sfienti}},
  \bibinfo {author} {\bibfnamefont {J.}~\bibnamefont {Piekarewicz}}, \bibinfo
  {author} {\bibfnamefont {C.~J.}\ \bibnamefont {Horowitz}}, \ and\ \bibinfo
  {author} {\bibfnamefont {M.}~\bibnamefont {Vanderhaeghen}},\ }\href {\doibase
  10.1088/1361-6471/ab2c6d} {\bibfield  {journal} {\bibinfo  {journal} {J.
  Phys. G: Nucl. Part. Phys.}\ }\textbf {\bibinfo {volume} {46}},\ \bibinfo
  {pages} {093003} (\bibinfo {year} {2019})}\BibitemShut {NoStop}%
\bibitem [{\citenamefont {Reed}\ \emph {et~al.}(2021)\citenamefont {Reed},
  \citenamefont {Fattoyev}, \citenamefont {Horowitz},\ and\ \citenamefont
  {Piekarewicz}}]{Reed:2021nqk}%
  \BibitemOpen
  \bibfield  {author} {\bibinfo {author} {\bibfnamefont {B.~T.}\ \bibnamefont
  {Reed}}, \bibinfo {author} {\bibfnamefont {F.~J.}\ \bibnamefont {Fattoyev}},
  \bibinfo {author} {\bibfnamefont {C.~J.}\ \bibnamefont {Horowitz}}, \ and\
  \bibinfo {author} {\bibfnamefont {J.}~\bibnamefont {Piekarewicz}},\ }\href
  {\doibase 10.1103/PhysRevLett.126.172503} {\bibfield  {journal} {\bibinfo
  {journal} {Phys. Rev. Lett.}\ }\textbf {\bibinfo {volume} {126}},\ \bibinfo
  {pages} {172503} (\bibinfo {year} {2021})},\ \Eprint
  {http://arxiv.org/abs/2101.03193} {arXiv:2101.03193 [nucl-th]} \BibitemShut
  {NoStop}%
\bibitem [{\citenamefont {Trzci\'{n}ska}\ \emph {et~al.}(2001)\citenamefont
  {Trzci\'{n}ska}, \citenamefont {Jastrz{\c{e}}bski}, \citenamefont
  {Lubi\'{n}ski}, \citenamefont {Hartmann}, \citenamefont {Schmidt},
  \citenamefont {von Egidy},\ and\ \citenamefont {K{\l}os}}]{PRL.87.082501}%
  \BibitemOpen
  \bibfield  {author} {\bibinfo {author} {\bibfnamefont {A.}~\bibnamefont
  {Trzci\'{n}ska}}, \bibinfo {author} {\bibfnamefont {J.}~\bibnamefont
  {Jastrz{\c{e}}bski}}, \bibinfo {author} {\bibfnamefont {P.}~\bibnamefont
  {Lubi\'{n}ski}}, \bibinfo {author} {\bibfnamefont {F.~J.}\ \bibnamefont
  {Hartmann}}, \bibinfo {author} {\bibfnamefont {R.}~\bibnamefont {Schmidt}},
  \bibinfo {author} {\bibfnamefont {T.}~\bibnamefont {von Egidy}}, \ and\
  \bibinfo {author} {\bibfnamefont {B.}~\bibnamefont {K{\l}os}},\ }\href
  {\doibase 10.1103/PhysRevLett.87.082501} {\bibfield  {journal} {\bibinfo
  {journal} {Phys. Rev. Lett.}\ }\textbf {\bibinfo {volume} {87}},\ \bibinfo
  {pages} {082501} (\bibinfo {year} {2001})}\BibitemShut {NoStop}%
\bibitem [{\citenamefont {Zenihiro}\ \emph {et~al.}(2010)\citenamefont
  {Zenihiro}, \citenamefont {Sakaguchi}, \citenamefont {Murakami},
  \citenamefont {Yosoi}, \citenamefont {Yasuda}, \citenamefont {Terashima},
  \citenamefont {Iwao} \emph {et~al.}}]{PRC.82.044611}%
  \BibitemOpen
  \bibfield  {author} {\bibinfo {author} {\bibfnamefont {J.}~\bibnamefont
  {Zenihiro}}, \bibinfo {author} {\bibfnamefont {H.}~\bibnamefont {Sakaguchi}},
  \bibinfo {author} {\bibfnamefont {T.}~\bibnamefont {Murakami}}, \bibinfo
  {author} {\bibfnamefont {M.}~\bibnamefont {Yosoi}}, \bibinfo {author}
  {\bibfnamefont {Y.}~\bibnamefont {Yasuda}}, \bibinfo {author} {\bibfnamefont
  {S.}~\bibnamefont {Terashima}}, \bibinfo {author} {\bibfnamefont
  {Y.}~\bibnamefont {Iwao}},  \emph {et~al.},\ }\href {\doibase
  10.1103/PhysRevC.82.044611} {\bibfield  {journal} {\bibinfo  {journal} {Phys.
  Rev. C}\ }\textbf {\bibinfo {volume} {82}},\ \bibinfo {pages} {044611}
  (\bibinfo {year} {2010})}\BibitemShut {NoStop}%
\bibitem [{\citenamefont {Tamii}\ \emph
  {et~al.}(2011{\natexlab{b}})\citenamefont {Tamii}, \citenamefont
  {Poltoratska}, \citenamefont {von Neumann-Cosel}, \citenamefont {Fujita},
  \citenamefont {Adachi}, \citenamefont {Bertulani}, \citenamefont {Carter}
  \emph {et~al.}}]{PRL.107.062502}%
  \BibitemOpen
  \bibfield  {author} {\bibinfo {author} {\bibfnamefont {A.}~\bibnamefont
  {Tamii}}, \bibinfo {author} {\bibfnamefont {I.}~\bibnamefont {Poltoratska}},
  \bibinfo {author} {\bibfnamefont {P.}~\bibnamefont {von Neumann-Cosel}},
  \bibinfo {author} {\bibfnamefont {Y.}~\bibnamefont {Fujita}}, \bibinfo
  {author} {\bibfnamefont {T.}~\bibnamefont {Adachi}}, \bibinfo {author}
  {\bibfnamefont {C.~A.}\ \bibnamefont {Bertulani}}, \bibinfo {author}
  {\bibfnamefont {J.}~\bibnamefont {Carter}},  \emph {et~al.},\ }\href
  {\doibase 10.1103/PhysRevLett.107.062502} {\bibfield  {journal} {\bibinfo
  {journal} {Phys. Rev. Lett.}\ }\textbf {\bibinfo {volume} {107}},\ \bibinfo
  {pages} {062502} (\bibinfo {year} {2011}{\natexlab{b}})}\BibitemShut
  {NoStop}%
\bibitem [{\citenamefont {Tarbert}\ \emph {et~al.}(2014)\citenamefont
  {Tarbert}, \citenamefont {Watts}, \citenamefont {Glazier}, \citenamefont
  {Aguar}, \citenamefont {Ahrens}, \citenamefont {Annand}, \citenamefont
  {Arends}, \citenamefont {Beck}, \citenamefont {Bekrenev}, \citenamefont
  {Boillat} \emph {et~al.}}]{PRL.112.242502}%
  \BibitemOpen
  \bibfield  {author} {\bibinfo {author} {\bibfnamefont {C.~M.}\ \bibnamefont
  {Tarbert}}, \bibinfo {author} {\bibfnamefont {D.~P.}\ \bibnamefont {Watts}},
  \bibinfo {author} {\bibfnamefont {D.~I.}\ \bibnamefont {Glazier}}, \bibinfo
  {author} {\bibfnamefont {P.}~\bibnamefont {Aguar}}, \bibinfo {author}
  {\bibfnamefont {J.}~\bibnamefont {Ahrens}}, \bibinfo {author} {\bibfnamefont
  {J.~R.~M.}\ \bibnamefont {Annand}}, \bibinfo {author} {\bibfnamefont {H.~J.}\
  \bibnamefont {Arends}}, \bibinfo {author} {\bibfnamefont {R.}~\bibnamefont
  {Beck}}, \bibinfo {author} {\bibfnamefont {V.}~\bibnamefont {Bekrenev}},
  \bibinfo {author} {\bibfnamefont {B.}~\bibnamefont {Boillat}},  \emph
  {et~al.} (\bibinfo {collaboration} {Crystal Ball at MAMI and A2
  Collaboration}),\ }\href {\doibase 10.1103/PhysRevLett.112.242502} {\bibfield
   {journal} {\bibinfo  {journal} {Phys. Rev. Lett.}\ }\textbf {\bibinfo
  {volume} {112}},\ \bibinfo {pages} {242502} (\bibinfo {year}
  {2014})}\BibitemShut {NoStop}%
\bibitem [{\citenamefont {Atkinson}\ \emph {et~al.}(2020)\citenamefont
  {Atkinson}, \citenamefont {Mahzoon}, \citenamefont {Keim}, \citenamefont
  {Bordelon}, \citenamefont {Pruitt}, \citenamefont {Charity},\ and\
  \citenamefont {Dickhoff}}]{PRC.101.044303}%
  \BibitemOpen
  \bibfield  {author} {\bibinfo {author} {\bibfnamefont {M.~C.}\ \bibnamefont
  {Atkinson}}, \bibinfo {author} {\bibfnamefont {M.~H.}\ \bibnamefont
  {Mahzoon}}, \bibinfo {author} {\bibfnamefont {M.~A.}\ \bibnamefont {Keim}},
  \bibinfo {author} {\bibfnamefont {B.~A.}\ \bibnamefont {Bordelon}}, \bibinfo
  {author} {\bibfnamefont {C.~D.}\ \bibnamefont {Pruitt}}, \bibinfo {author}
  {\bibfnamefont {R.~J.}\ \bibnamefont {Charity}}, \ and\ \bibinfo {author}
  {\bibfnamefont {W.~H.}\ \bibnamefont {Dickhoff}},\ }\href {\doibase
  10.1103/PhysRevC.101.044303} {\bibfield  {journal} {\bibinfo  {journal}
  {Phys. Rev. C}\ }\textbf {\bibinfo {volume} {101}},\ \bibinfo {pages}
  {044303} (\bibinfo {year} {2020})}\BibitemShut {NoStop}%
\bibitem [{\citenamefont {Hagen}\ \emph {et~al.}(2014)\citenamefont {Hagen},
  \citenamefont {Papenbrock}, \citenamefont {Hjorth-Jensen},\ and\
  \citenamefont {Dean}}]{Hagen:2013nca}%
  \BibitemOpen
  \bibfield  {author} {\bibinfo {author} {\bibfnamefont {G.}~\bibnamefont
  {Hagen}}, \bibinfo {author} {\bibfnamefont {T.}~\bibnamefont {Papenbrock}},
  \bibinfo {author} {\bibfnamefont {M.}~\bibnamefont {Hjorth-Jensen}}, \ and\
  \bibinfo {author} {\bibfnamefont {D.~J.}\ \bibnamefont {Dean}},\ }\href
  {\doibase 10.1088/0034-4885/77/9/096302} {\bibfield  {journal} {\bibinfo
  {journal} {Rept. Prog. Phys.}\ }\textbf {\bibinfo {volume} {77}},\ \bibinfo
  {pages} {096302} (\bibinfo {year} {2014})},\ \Eprint
  {http://arxiv.org/abs/1312.7872} {arXiv:1312.7872 [nucl-th]} \BibitemShut
  {NoStop}%
\bibitem [{\citenamefont {Hagen}\ \emph {et~al.}(2015)\citenamefont {Hagen}
  \emph {et~al.}}]{Hagen:2015yea}%
  \BibitemOpen
  \bibfield  {author} {\bibinfo {author} {\bibfnamefont {G.}~\bibnamefont
  {Hagen}} \emph {et~al.},\ }\href {\doibase 10.1038/nphys3529} {\bibfield
  {journal} {\bibinfo  {journal} {Nature Phys.}\ }\textbf {\bibinfo {volume}
  {12}},\ \bibinfo {pages} {186} (\bibinfo {year} {2015})},\ \Eprint
  {http://arxiv.org/abs/1509.07169} {arXiv:1509.07169 [nucl-th]} \BibitemShut
  {NoStop}%
\bibitem [{\citenamefont {Kohno}(2012)}]{Kohno:2012vj}%
  \BibitemOpen
  \bibfield  {author} {\bibinfo {author} {\bibfnamefont {M.}~\bibnamefont
  {Kohno}},\ }\href {\doibase 10.1103/PhysRevC.86.061301} {\bibfield  {journal}
  {\bibinfo  {journal} {Phys. Rev. C}\ }\textbf {\bibinfo {volume} {86}},\
  \bibinfo {pages} {061301} (\bibinfo {year} {2012})},\ \Eprint
  {http://arxiv.org/abs/1209.5048} {arXiv:1209.5048 [nucl-th]} \BibitemShut
  {NoStop}%
\bibitem [{\citenamefont {Toyokawa}\ \emph {et~al.}(2018)\citenamefont
  {Toyokawa}, \citenamefont {Yahiro}, \citenamefont {Matsumoto},\ and\
  \citenamefont {Kohno}}]{Toyokawa:2017pdd}%
  \BibitemOpen
  \bibfield  {author} {\bibinfo {author} {\bibfnamefont {M.}~\bibnamefont
  {Toyokawa}}, \bibinfo {author} {\bibfnamefont {M.}~\bibnamefont {Yahiro}},
  \bibinfo {author} {\bibfnamefont {T.}~\bibnamefont {Matsumoto}}, \ and\
  \bibinfo {author} {\bibfnamefont {M.}~\bibnamefont {Kohno}},\ }\href
  {\doibase 10.1093/ptep/pty001} {\bibfield  {journal} {\bibinfo  {journal}
  {PTEP}\ }\textbf {\bibinfo {volume} {2018}},\ \bibinfo {pages} {023D03}
  (\bibinfo {year} {2018})},\ \Eprint {http://arxiv.org/abs/1712.07033}
  {arXiv:1712.07033 [nucl-th]} \BibitemShut {NoStop}%
\bibitem [{\citenamefont {von Geramb}\ \emph {et~al.}(1991)\citenamefont {von
  Geramb} \emph {et~al.}}]{von-Geramb-1991}%
  \BibitemOpen
  \bibfield  {author} {\bibinfo {author} {\bibfnamefont {H.~V.}\ \bibnamefont
  {von Geramb}} \emph {et~al.},\ }\href@noop {} {\bibfield  {journal} {\bibinfo
   {journal} {Phys. Rev. C}\ }\textbf {\bibinfo {volume} {44}},\ \bibinfo
  {pages} {73} (\bibinfo {year} {1991})}\BibitemShut {NoStop}%
\bibitem [{\citenamefont {Amos}\ and\ \citenamefont
  {Dortmans}(1994)}]{Amos-1994}%
  \BibitemOpen
  \bibfield  {author} {\bibinfo {author} {\bibfnamefont {K.}~\bibnamefont
  {Amos}}\ and\ \bibinfo {author} {\bibfnamefont {P.~J.}\ \bibnamefont
  {Dortmans}},\ }\href@noop {} {\bibfield  {journal} {\bibinfo  {journal}
  {Phys. Rev. C}\ }\textbf {\bibinfo {volume} {49}},\ \bibinfo {pages} {1309}
  (\bibinfo {year} {1994})}\BibitemShut {NoStop}%
\bibitem [{\citenamefont {Tagami}\ \emph {et~al.}(2020)\citenamefont {Tagami},
  \citenamefont {Tanaka}, \citenamefont {Takechi}, \citenamefont {Fukuda},\
  and\ \citenamefont {Yahiro}}]{Tagami:2019svt}%
  \BibitemOpen
  \bibfield  {author} {\bibinfo {author} {\bibfnamefont {S.}~\bibnamefont
  {Tagami}}, \bibinfo {author} {\bibfnamefont {M.}~\bibnamefont {Tanaka}},
  \bibinfo {author} {\bibfnamefont {M.}~\bibnamefont {Takechi}}, \bibinfo
  {author} {\bibfnamefont {M.}~\bibnamefont {Fukuda}}, \ and\ \bibinfo {author}
  {\bibfnamefont {M.}~\bibnamefont {Yahiro}},\ }\href {\doibase
  10.1103/PhysRevC.101.014620} {\bibfield  {journal} {\bibinfo  {journal}
  {Phys. Rev. C}\ }\textbf {\bibinfo {volume} {101}},\ \bibinfo {pages}
  {014620} (\bibinfo {year} {2020})},\ \Eprint
  {http://arxiv.org/abs/1911.05417} {arXiv:1911.05417 [nucl-th]} \BibitemShut
  {NoStop}%
\bibitem [{\citenamefont {Toyokawa}\ \emph
  {et~al.}(2015{\natexlab{a}})\citenamefont {Toyokawa}, \citenamefont {Minomo},
  \citenamefont {Kohno},\ and\ \citenamefont {Yahiro}}]{Toyokawa:2014yma}%
  \BibitemOpen
  \bibfield  {author} {\bibinfo {author} {\bibfnamefont {M.}~\bibnamefont
  {Toyokawa}}, \bibinfo {author} {\bibfnamefont {K.}~\bibnamefont {Minomo}},
  \bibinfo {author} {\bibfnamefont {M.}~\bibnamefont {Kohno}}, \ and\ \bibinfo
  {author} {\bibfnamefont {M.}~\bibnamefont {Yahiro}},\ }\href {\doibase
  10.1088/0954-3899/42/2/025104} {\bibfield  {journal} {\bibinfo  {journal} {J.
  Phys. G}\ }\textbf {\bibinfo {volume} {42}},\ \bibinfo {pages} {025104}
  (\bibinfo {year} {2015}{\natexlab{a}})},\ \bibinfo {note} {[Erratum: J.Phys.G
  44, 079502 (2017)]},\ \Eprint {http://arxiv.org/abs/1404.6895}
  {arXiv:1404.6895 [nucl-th]} \BibitemShut {NoStop}%
\bibitem [{\citenamefont {Toyokawa}\ \emph
  {et~al.}(2015{\natexlab{b}})\citenamefont {Toyokawa}, \citenamefont {Yahiro},
  \citenamefont {Matsumoto}, \citenamefont {Minomo}, \citenamefont {Ogata},\
  and\ \citenamefont {Kohno}}]{Toyokawa:2015zxa}%
  \BibitemOpen
  \bibfield  {author} {\bibinfo {author} {\bibfnamefont {M.}~\bibnamefont
  {Toyokawa}}, \bibinfo {author} {\bibfnamefont {M.}~\bibnamefont {Yahiro}},
  \bibinfo {author} {\bibfnamefont {T.}~\bibnamefont {Matsumoto}}, \bibinfo
  {author} {\bibfnamefont {K.}~\bibnamefont {Minomo}}, \bibinfo {author}
  {\bibfnamefont {K.}~\bibnamefont {Ogata}}, \ and\ \bibinfo {author}
  {\bibfnamefont {M.}~\bibnamefont {Kohno}},\ }\href {\doibase
  10.1103/PhysRevC.92.024618} {\bibfield  {journal} {\bibinfo  {journal} {Phys.
  Rev. C}\ }\textbf {\bibinfo {volume} {92}},\ \bibinfo {pages} {024618}
  (\bibinfo {year} {2015}{\natexlab{b}})},\ \bibinfo {note} {[Erratum:
  Phys.Rev.C 96, 059905 (2017)]},\ \Eprint {http://arxiv.org/abs/1507.02807}
  {arXiv:1507.02807 [nucl-th]} \BibitemShut {NoStop}%
\bibitem [{\citenamefont {Tagami}\ \emph {et~al.}(2021)\citenamefont {Tagami},
  \citenamefont {Wakasa}, \citenamefont {Matsui}, \citenamefont {Yahiro},\ and\
  \citenamefont {Takechi}}]{Tagami:2020bee}%
  \BibitemOpen
  \bibfield  {author} {\bibinfo {author} {\bibfnamefont {S.}~\bibnamefont
  {Tagami}}, \bibinfo {author} {\bibfnamefont {T.}~\bibnamefont {Wakasa}},
  \bibinfo {author} {\bibfnamefont {J.}~\bibnamefont {Matsui}}, \bibinfo
  {author} {\bibfnamefont {M.}~\bibnamefont {Yahiro}}, \ and\ \bibinfo {author}
  {\bibfnamefont {M.}~\bibnamefont {Takechi}},\ }\href {\doibase
  10.1103/PhysRevC.104.024606} {\bibfield  {journal} {\bibinfo  {journal}
  {Phys. Rev. C}\ }\textbf {\bibinfo {volume} {104}},\ \bibinfo {pages}
  {024606} (\bibinfo {year} {2021})},\ \Eprint
  {http://arxiv.org/abs/2010.02450} {arXiv:2010.02450 [nucl-th]} \BibitemShut
  {NoStop}%
\bibitem [{\citenamefont {Roca-Maza}\ \emph {et~al.}(2011)\citenamefont
  {Roca-Maza}, \citenamefont {Centelles}, \citenamefont {Vinas},\ and\
  \citenamefont {Warda}}]{RocaMaza:2011pm}%
  \BibitemOpen
  \bibfield  {author} {\bibinfo {author} {\bibfnamefont {X.}~\bibnamefont
  {Roca-Maza}}, \bibinfo {author} {\bibfnamefont {M.}~\bibnamefont
  {Centelles}}, \bibinfo {author} {\bibfnamefont {X.}~\bibnamefont {Vinas}}, \
  and\ \bibinfo {author} {\bibfnamefont {M.}~\bibnamefont {Warda}},\ }\href
  {\doibase 10.1103/PhysRevLett.106.252501} {\bibfield  {journal} {\bibinfo
  {journal} {Phys. Rev. Lett.}\ }\textbf {\bibinfo {volume} {106}},\ \bibinfo
  {pages} {252501} (\bibinfo {year} {2011})},\ \Eprint
  {http://arxiv.org/abs/1103.1762} {arXiv:1103.1762 [nucl-th]} \BibitemShut
  {NoStop}%
\bibitem [{\citenamefont {Akmal}\ \emph {et~al.}(1998)\citenamefont {Akmal},
  \citenamefont {Pandharipande},\ and\ \citenamefont
  {Ravenhall}}]{Akmal:1998cf}%
  \BibitemOpen
  \bibfield  {author} {\bibinfo {author} {\bibfnamefont {A.}~\bibnamefont
  {Akmal}}, \bibinfo {author} {\bibfnamefont {V.~R.}\ \bibnamefont
  {Pandharipande}}, \ and\ \bibinfo {author} {\bibfnamefont {D.~G.}\
  \bibnamefont {Ravenhall}},\ }\href {\doibase 10.1103/PhysRevC.58.1804}
  {\bibfield  {journal} {\bibinfo  {journal} {Phys. Rev. C}\ }\textbf {\bibinfo
  {volume} {58}},\ \bibinfo {pages} {1804} (\bibinfo {year} {1998})},\ \Eprint
  {http://arxiv.org/abs/nucl-th/9804027} {arXiv:nucl-th/9804027} \BibitemShut
  {NoStop}%
\bibitem [{\citenamefont {Ishizuka}\ \emph {et~al.}(2015)\citenamefont
  {Ishizuka}, \citenamefont {Suda}, \citenamefont {Suzuki}, \citenamefont
  {Ohnishi}, \citenamefont {Sumiyoshi},\ and\ \citenamefont
  {Toki}}]{Ishizuka:2014jsa}%
  \BibitemOpen
  \bibfield  {author} {\bibinfo {author} {\bibfnamefont {C.}~\bibnamefont
  {Ishizuka}}, \bibinfo {author} {\bibfnamefont {T.}~\bibnamefont {Suda}},
  \bibinfo {author} {\bibfnamefont {H.}~\bibnamefont {Suzuki}}, \bibinfo
  {author} {\bibfnamefont {A.}~\bibnamefont {Ohnishi}}, \bibinfo {author}
  {\bibfnamefont {K.}~\bibnamefont {Sumiyoshi}}, \ and\ \bibinfo {author}
  {\bibfnamefont {H.}~\bibnamefont {Toki}},\ }\href {\doibase
  10.1093/pasj/psu141} {\bibfield  {journal} {\bibinfo  {journal} {Publ.
  Astron. Soc. Jap.}\ }\textbf {\bibinfo {volume} {67}},\ \bibinfo {pages} {13}
  (\bibinfo {year} {2015})},\ \Eprint {http://arxiv.org/abs/1408.6230}
  {arXiv:1408.6230 [nucl-th]} \BibitemShut {NoStop}%
\bibitem [{\citenamefont {Gonzalez-Boquera}\ \emph {et~al.}(2018)\citenamefont
  {Gonzalez-Boquera}, \citenamefont {Centelles}, \citenamefont {Vi\~nas},\ and\
  \citenamefont {Robledo}}]{Gonzalez-Boquera:2017rzy}%
  \BibitemOpen
  \bibfield  {author} {\bibinfo {author} {\bibfnamefont {C.}~\bibnamefont
  {Gonzalez-Boquera}}, \bibinfo {author} {\bibfnamefont {M.}~\bibnamefont
  {Centelles}}, \bibinfo {author} {\bibfnamefont {X.}~\bibnamefont {Vi\~nas}},
  \ and\ \bibinfo {author} {\bibfnamefont {L.~M.}\ \bibnamefont {Robledo}},\
  }\href {\doibase 10.1016/j.physletb.2018.02.005} {\bibfield  {journal}
  {\bibinfo  {journal} {Phys. Lett. B}\ }\textbf {\bibinfo {volume} {779}},\
  \bibinfo {pages} {195} (\bibinfo {year} {2018})},\ \Eprint
  {http://arxiv.org/abs/1712.06735} {arXiv:1712.06735 [nucl-th]} \BibitemShut
  {NoStop}%
\bibitem [{\citenamefont {Farine}\ \emph {et~al.}(1999)\citenamefont {Farine},
  \citenamefont {Von-Eiff}, \citenamefont {Schuck}, \citenamefont {Berger},
  \citenamefont {Decharg{\'{e}}},\ and\ \citenamefont {Girod}}]{D1P-1999}%
  \BibitemOpen
  \bibfield  {author} {\bibinfo {author} {\bibfnamefont {M.}~\bibnamefont
  {Farine}}, \bibinfo {author} {\bibfnamefont {D.}~\bibnamefont {Von-Eiff}},
  \bibinfo {author} {\bibfnamefont {P.}~\bibnamefont {Schuck}}, \bibinfo
  {author} {\bibfnamefont {J.~F.}\ \bibnamefont {Berger}}, \bibinfo {author}
  {\bibfnamefont {J.}~\bibnamefont {Decharg{\'{e}}}}, \ and\ \bibinfo {author}
  {\bibfnamefont {M.}~\bibnamefont {Girod}},\ }\href {\doibase
  10.1088/0954-3899/25/4/056} {\ \textbf {\bibinfo {volume} {25}},\ \bibinfo
  {pages} {863} (\bibinfo {year} {1999})}\BibitemShut {NoStop}%
\bibitem [{\citenamefont {Gonzalez-Boquera}\ \emph {et~al.}(2017)\citenamefont
  {Gonzalez-Boquera}, \citenamefont {Centelles}, \citenamefont {Vi\~nas},\ and\
  \citenamefont {Rios}}]{Gonzalez-Boquera:2017uep}%
  \BibitemOpen
  \bibfield  {author} {\bibinfo {author} {\bibfnamefont {C.}~\bibnamefont
  {Gonzalez-Boquera}}, \bibinfo {author} {\bibfnamefont {M.}~\bibnamefont
  {Centelles}}, \bibinfo {author} {\bibfnamefont {X.}~\bibnamefont {Vi\~nas}},
  \ and\ \bibinfo {author} {\bibfnamefont {A.}~\bibnamefont {Rios}},\ }\href
  {\doibase 10.1103/PhysRevC.96.065806} {\bibfield  {journal} {\bibinfo
  {journal} {Phys. Rev. C}\ }\textbf {\bibinfo {volume} {96}},\ \bibinfo
  {pages} {065806} (\bibinfo {year} {2017})},\ \Eprint
  {http://arxiv.org/abs/1706.02736} {arXiv:1706.02736 [nucl-th]} \BibitemShut
  {NoStop}%
\bibitem [{\citenamefont {Oertel}\ \emph {et~al.}(2017)\citenamefont {Oertel},
  \citenamefont {Hempel}, \citenamefont {Kl\"ahn},\ and\ \citenamefont
  {Typel}}]{Oertel:2016bki}%
  \BibitemOpen
  \bibfield  {author} {\bibinfo {author} {\bibfnamefont {M.}~\bibnamefont
  {Oertel}}, \bibinfo {author} {\bibfnamefont {M.}~\bibnamefont {Hempel}},
  \bibinfo {author} {\bibfnamefont {T.}~\bibnamefont {Kl\"ahn}}, \ and\
  \bibinfo {author} {\bibfnamefont {S.}~\bibnamefont {Typel}},\ }\href
  {\doibase 10.1103/RevModPhys.89.015007} {\bibfield  {journal} {\bibinfo
  {journal} {Rev. Mod. Phys.}\ }\textbf {\bibinfo {volume} {89}},\ \bibinfo
  {pages} {015007} (\bibinfo {year} {2017})},\ \Eprint
  {http://arxiv.org/abs/1610.03361} {arXiv:1610.03361 [astro-ph.HE]}
  \BibitemShut {NoStop}%
\bibitem [{\citenamefont {Piekarewicz}(2007)}]{Piekarewicz:2007dx}%
  \BibitemOpen
  \bibfield  {author} {\bibinfo {author} {\bibfnamefont {J.}~\bibnamefont
  {Piekarewicz}},\ }\href {\doibase 10.1103/PhysRevC.76.064310} {\bibfield
  {journal} {\bibinfo  {journal} {Phys. Rev. C}\ }\textbf {\bibinfo {volume}
  {76}},\ \bibinfo {pages} {064310} (\bibinfo {year} {2007})},\ \Eprint
  {http://arxiv.org/abs/0709.2699} {arXiv:0709.2699 [nucl-th]} \BibitemShut
  {NoStop}%
\bibitem [{\citenamefont {Lim}\ \emph {et~al.}(2014)\citenamefont {Lim},
  \citenamefont {Kwak}, \citenamefont {Hyun},\ and\ \citenamefont
  {Lee}}]{Lim:2013tqa}%
  \BibitemOpen
  \bibfield  {author} {\bibinfo {author} {\bibfnamefont {Y.}~\bibnamefont
  {Lim}}, \bibinfo {author} {\bibfnamefont {K.}~\bibnamefont {Kwak}}, \bibinfo
  {author} {\bibfnamefont {C.~H.}\ \bibnamefont {Hyun}}, \ and\ \bibinfo
  {author} {\bibfnamefont {C.-H.}\ \bibnamefont {Lee}},\ }\href {\doibase
  10.1103/PhysRevC.89.055804} {\bibfield  {journal} {\bibinfo  {journal} {Phys.
  Rev. C}\ }\textbf {\bibinfo {volume} {89}},\ \bibinfo {pages} {055804}
  (\bibinfo {year} {2014})},\ \Eprint {http://arxiv.org/abs/1312.2640}
  {arXiv:1312.2640 [nucl-th]} \BibitemShut {NoStop}%
\bibitem [{\citenamefont {Sellahewa}\ and\ \citenamefont
  {Rios}(2014)}]{Sellahewa:2014nia}%
  \BibitemOpen
  \bibfield  {author} {\bibinfo {author} {\bibfnamefont {R.}~\bibnamefont
  {Sellahewa}}\ and\ \bibinfo {author} {\bibfnamefont {A.}~\bibnamefont
  {Rios}},\ }\href {\doibase 10.1103/PhysRevC.90.054327} {\bibfield  {journal}
  {\bibinfo  {journal} {Phys. Rev. C}\ }\textbf {\bibinfo {volume} {90}},\
  \bibinfo {pages} {054327} (\bibinfo {year} {2014})},\ \Eprint
  {http://arxiv.org/abs/1407.8138} {arXiv:1407.8138 [nucl-th]} \BibitemShut
  {NoStop}%
\bibitem [{\citenamefont {Inakura}\ and\ \citenamefont
  {Nakada}(2015)}]{Inakura:2015cla}%
  \BibitemOpen
  \bibfield  {author} {\bibinfo {author} {\bibfnamefont {T.}~\bibnamefont
  {Inakura}}\ and\ \bibinfo {author} {\bibfnamefont {H.}~\bibnamefont
  {Nakada}},\ }\href {\doibase 10.1103/PhysRevC.92.064302} {\bibfield
  {journal} {\bibinfo  {journal} {Phys. Rev. C}\ }\textbf {\bibinfo {volume}
  {92}},\ \bibinfo {pages} {064302} (\bibinfo {year} {2015})},\ \Eprint
  {http://arxiv.org/abs/1509.02982} {arXiv:1509.02982 [nucl-th]} \BibitemShut
  {NoStop}%
\bibitem [{\citenamefont {Fattoyev}\ and\ \citenamefont
  {Piekarewicz}(2013)}]{Fattoyev:2013yaa}%
  \BibitemOpen
  \bibfield  {author} {\bibinfo {author} {\bibfnamefont {F.~J.}\ \bibnamefont
  {Fattoyev}}\ and\ \bibinfo {author} {\bibfnamefont {J.}~\bibnamefont
  {Piekarewicz}},\ }\href {\doibase 10.1103/PhysRevLett.111.162501} {\bibfield
  {journal} {\bibinfo  {journal} {Phys. Rev. Lett.}\ }\textbf {\bibinfo
  {volume} {111}},\ \bibinfo {pages} {162501} (\bibinfo {year} {2013})},\
  \Eprint {http://arxiv.org/abs/1306.6034} {arXiv:1306.6034 [nucl-th]}
  \BibitemShut {NoStop}%
\bibitem [{\citenamefont {Steiner}\ \emph {et~al.}(2005)\citenamefont
  {Steiner}, \citenamefont {Prakash}, \citenamefont {Lattimer},\ and\
  \citenamefont {Ellis}}]{Steiner:2004fi}%
  \BibitemOpen
  \bibfield  {author} {\bibinfo {author} {\bibfnamefont {A.~W.}\ \bibnamefont
  {Steiner}}, \bibinfo {author} {\bibfnamefont {M.}~\bibnamefont {Prakash}},
  \bibinfo {author} {\bibfnamefont {J.~M.}\ \bibnamefont {Lattimer}}, \ and\
  \bibinfo {author} {\bibfnamefont {P.~J.}\ \bibnamefont {Ellis}},\ }\href
  {\doibase 10.1016/j.physrep.2005.02.004} {\bibfield  {journal} {\bibinfo
  {journal} {Phys. Rept.}\ }\textbf {\bibinfo {volume} {411}},\ \bibinfo
  {pages} {325} (\bibinfo {year} {2005})},\ \Eprint
  {http://arxiv.org/abs/nucl-th/0410066} {arXiv:nucl-th/0410066} \BibitemShut
  {NoStop}%
\bibitem [{\citenamefont {Centelles}\ \emph {et~al.}(2010)\citenamefont
  {Centelles}, \citenamefont {Roca-Maza}, \citenamefont {Vinas},\ and\
  \citenamefont {Warda}}]{Centelles:2010qh}%
  \BibitemOpen
  \bibfield  {author} {\bibinfo {author} {\bibfnamefont {M.}~\bibnamefont
  {Centelles}}, \bibinfo {author} {\bibfnamefont {X.}~\bibnamefont
  {Roca-Maza}}, \bibinfo {author} {\bibfnamefont {X.}~\bibnamefont {Vinas}}, \
  and\ \bibinfo {author} {\bibfnamefont {M.}~\bibnamefont {Warda}},\ }\href
  {\doibase 10.1103/PhysRevC.82.054314} {\bibfield  {journal} {\bibinfo
  {journal} {Phys. Rev. C}\ }\textbf {\bibinfo {volume} {82}},\ \bibinfo
  {pages} {054314} (\bibinfo {year} {2010})},\ \Eprint
  {http://arxiv.org/abs/1010.5396} {arXiv:1010.5396 [nucl-th]} \BibitemShut
  {NoStop}%
\bibitem [{\citenamefont {Dutra}\ \emph {et~al.}(2012)\citenamefont {Dutra},
  \citenamefont {Lourenco}, \citenamefont {Sa~Martins}, \citenamefont
  {Delfino}, \citenamefont {Stone},\ and\ \citenamefont
  {Stevenson}}]{Dutra:2012mb}%
  \BibitemOpen
  \bibfield  {author} {\bibinfo {author} {\bibfnamefont {M.}~\bibnamefont
  {Dutra}}, \bibinfo {author} {\bibfnamefont {O.}~\bibnamefont {Lourenco}},
  \bibinfo {author} {\bibfnamefont {J.~S.}\ \bibnamefont {Sa~Martins}},
  \bibinfo {author} {\bibfnamefont {A.}~\bibnamefont {Delfino}}, \bibinfo
  {author} {\bibfnamefont {J.~R.}\ \bibnamefont {Stone}}, \ and\ \bibinfo
  {author} {\bibfnamefont {P.~D.}\ \bibnamefont {Stevenson}},\ }\href {\doibase
  10.1103/PhysRevC.85.035201} {\bibfield  {journal} {\bibinfo  {journal} {Phys.
  Rev. C}\ }\textbf {\bibinfo {volume} {85}},\ \bibinfo {pages} {035201}
  (\bibinfo {year} {2012})},\ \Eprint {http://arxiv.org/abs/1202.3902}
  {arXiv:1202.3902 [nucl-th]} \BibitemShut {NoStop}%
\bibitem [{\citenamefont {Brown}\ and\ \citenamefont
  {Schwenk}(2014)}]{Brown:2013pwa}%
  \BibitemOpen
  \bibfield  {author} {\bibinfo {author} {\bibfnamefont {B.~A.}\ \bibnamefont
  {Brown}}\ and\ \bibinfo {author} {\bibfnamefont {A.}~\bibnamefont
  {Schwenk}},\ }\href {\doibase 10.1103/PhysRevC.89.011307} {\bibfield
  {journal} {\bibinfo  {journal} {Phys. Rev. C}\ }\textbf {\bibinfo {volume}
  {89}},\ \bibinfo {pages} {011307} (\bibinfo {year} {2014})},\ \bibinfo {note}
  {[Erratum: Phys.Rev.C 91, 049902 (2015)]},\ \Eprint
  {http://arxiv.org/abs/1311.3957} {arXiv:1311.3957 [nucl-th]} \BibitemShut
  {NoStop}%
\bibitem [{\citenamefont {Brown}(2000)}]{Brown:2000pd}%
  \BibitemOpen
  \bibfield  {author} {\bibinfo {author} {\bibfnamefont {B.~A.}\ \bibnamefont
  {Brown}},\ }\href {\doibase 10.1103/PhysRevLett.85.5296} {\bibfield
  {journal} {\bibinfo  {journal} {Phys. Rev. Lett.}\ }\textbf {\bibinfo
  {volume} {85}},\ \bibinfo {pages} {5296} (\bibinfo {year}
  {2000})}\BibitemShut {NoStop}%
\bibitem [{\citenamefont {Reinhard}\ \emph {et~al.}(2016)\citenamefont
  {Reinhard}, \citenamefont {Umar}, \citenamefont {Stevenson}, \citenamefont
  {Piekarewicz}, \citenamefont {Oberacker},\ and\ \citenamefont
  {Maruhn}}]{Reinhard:2016sce}%
  \BibitemOpen
  \bibfield  {author} {\bibinfo {author} {\bibfnamefont {P.~G.}\ \bibnamefont
  {Reinhard}}, \bibinfo {author} {\bibfnamefont {A.~S.}\ \bibnamefont {Umar}},
  \bibinfo {author} {\bibfnamefont {P.~D.}\ \bibnamefont {Stevenson}}, \bibinfo
  {author} {\bibfnamefont {J.}~\bibnamefont {Piekarewicz}}, \bibinfo {author}
  {\bibfnamefont {V.~E.}\ \bibnamefont {Oberacker}}, \ and\ \bibinfo {author}
  {\bibfnamefont {J.~A.}\ \bibnamefont {Maruhn}},\ }\href {\doibase
  10.1103/PhysRevC.93.044618} {\bibfield  {journal} {\bibinfo  {journal} {Phys.
  Rev. C}\ }\textbf {\bibinfo {volume} {93}},\ \bibinfo {pages} {044618}
  (\bibinfo {year} {2016})},\ \Eprint {http://arxiv.org/abs/1603.01319}
  {arXiv:1603.01319 [nucl-th]} \BibitemShut {NoStop}%
\bibitem [{\citenamefont {Tsang}\ \emph {et~al.}(2019)\citenamefont {Tsang},
  \citenamefont {Brown}, \citenamefont {Fattoyev}, \citenamefont {Lynch},\ and\
  \citenamefont {Tsang}}]{Tsang:2019ymt}%
  \BibitemOpen
  \bibfield  {author} {\bibinfo {author} {\bibfnamefont {C.~Y.}\ \bibnamefont
  {Tsang}}, \bibinfo {author} {\bibfnamefont {B.~A.}\ \bibnamefont {Brown}},
  \bibinfo {author} {\bibfnamefont {F.~J.}\ \bibnamefont {Fattoyev}}, \bibinfo
  {author} {\bibfnamefont {W.~G.}\ \bibnamefont {Lynch}}, \ and\ \bibinfo
  {author} {\bibfnamefont {M.~B.}\ \bibnamefont {Tsang}},\ }\href {\doibase
  10.1103/PhysRevC.100.062801} {\bibfield  {journal} {\bibinfo  {journal}
  {Phys. Rev. C}\ }\textbf {\bibinfo {volume} {100}},\ \bibinfo {pages}
  {062801} (\bibinfo {year} {2019})},\ \Eprint
  {http://arxiv.org/abs/1908.11842} {arXiv:1908.11842 [nucl-th]} \BibitemShut
  {NoStop}%
\bibitem [{\citenamefont {Ducoin}\ \emph {et~al.}(2010)\citenamefont {Ducoin},
  \citenamefont {Margueron},\ and\ \citenamefont
  {Providencia}}]{Ducoin:2010as}%
  \BibitemOpen
  \bibfield  {author} {\bibinfo {author} {\bibfnamefont {C.}~\bibnamefont
  {Ducoin}}, \bibinfo {author} {\bibfnamefont {J.}~\bibnamefont {Margueron}}, \
  and\ \bibinfo {author} {\bibfnamefont {C.}~\bibnamefont {Providencia}},\
  }\href {\doibase 10.1209/0295-5075/91/32001} {\bibfield  {journal} {\bibinfo
  {journal} {EPL}\ }\textbf {\bibinfo {volume} {91}},\ \bibinfo {pages} {32001}
  (\bibinfo {year} {2010})},\ \Eprint {http://arxiv.org/abs/1004.5197}
  {arXiv:1004.5197 [nucl-th]} \BibitemShut {NoStop}%
\bibitem [{\citenamefont {Fortin}\ \emph {et~al.}(2016)\citenamefont {Fortin},
  \citenamefont {Providencia}, \citenamefont {Raduta}, \citenamefont
  {Gulminelli}, \citenamefont {Zdunik}, \citenamefont {Haensel},\ and\
  \citenamefont {Bejger}}]{Fortin:2016hny}%
  \BibitemOpen
  \bibfield  {author} {\bibinfo {author} {\bibfnamefont {M.}~\bibnamefont
  {Fortin}}, \bibinfo {author} {\bibfnamefont {C.}~\bibnamefont {Providencia}},
  \bibinfo {author} {\bibfnamefont {A.~R.}\ \bibnamefont {Raduta}}, \bibinfo
  {author} {\bibfnamefont {F.}~\bibnamefont {Gulminelli}}, \bibinfo {author}
  {\bibfnamefont {J.~L.}\ \bibnamefont {Zdunik}}, \bibinfo {author}
  {\bibfnamefont {P.}~\bibnamefont {Haensel}}, \ and\ \bibinfo {author}
  {\bibfnamefont {M.}~\bibnamefont {Bejger}},\ }\href {\doibase
  10.1103/PhysRevC.94.035804} {\bibfield  {journal} {\bibinfo  {journal} {Phys.
  Rev. C}\ }\textbf {\bibinfo {volume} {94}},\ \bibinfo {pages} {035804}
  (\bibinfo {year} {2016})},\ \Eprint {http://arxiv.org/abs/1604.01944}
  {arXiv:1604.01944 [astro-ph.SR]} \BibitemShut {NoStop}%
\bibitem [{\citenamefont {Chen}\ \emph {et~al.}(2010)\citenamefont {Chen},
  \citenamefont {Ko}, \citenamefont {Li},\ and\ \citenamefont
  {Xu}}]{Chen:2010qx}%
  \BibitemOpen
  \bibfield  {author} {\bibinfo {author} {\bibfnamefont {L.-W.}\ \bibnamefont
  {Chen}}, \bibinfo {author} {\bibfnamefont {C.~M.}\ \bibnamefont {Ko}},
  \bibinfo {author} {\bibfnamefont {B.-A.}\ \bibnamefont {Li}}, \ and\ \bibinfo
  {author} {\bibfnamefont {J.}~\bibnamefont {Xu}},\ }\href {\doibase
  10.1103/PhysRevC.82.024321} {\bibfield  {journal} {\bibinfo  {journal} {Phys.
  Rev. C}\ }\textbf {\bibinfo {volume} {82}},\ \bibinfo {pages} {024321}
  (\bibinfo {year} {2010})},\ \Eprint {http://arxiv.org/abs/1004.4672}
  {arXiv:1004.4672 [nucl-th]} \BibitemShut {NoStop}%
\bibitem [{\citenamefont {Zhao}\ and\ \citenamefont
  {Gandolfi}(2016)}]{Zhao:2016ujh}%
  \BibitemOpen
  \bibfield  {author} {\bibinfo {author} {\bibfnamefont {P.~W.}\ \bibnamefont
  {Zhao}}\ and\ \bibinfo {author} {\bibfnamefont {S.}~\bibnamefont
  {Gandolfi}},\ }\href {\doibase 10.1103/PhysRevC.94.041302} {\bibfield
  {journal} {\bibinfo  {journal} {Phys. Rev. C}\ }\textbf {\bibinfo {volume}
  {94}},\ \bibinfo {pages} {041302} (\bibinfo {year} {2016})},\ \Eprint
  {http://arxiv.org/abs/1604.01490} {arXiv:1604.01490 [nucl-th]} \BibitemShut
  {NoStop}%
\bibitem [{\citenamefont {Zhang}\ \emph {et~al.}(2018)\citenamefont {Zhang},
  \citenamefont {Lim}, \citenamefont {Holt},\ and\ \citenamefont
  {Ko}}]{Zhang:2017hvh}%
  \BibitemOpen
  \bibfield  {author} {\bibinfo {author} {\bibfnamefont {Z.}~\bibnamefont
  {Zhang}}, \bibinfo {author} {\bibfnamefont {Y.}~\bibnamefont {Lim}}, \bibinfo
  {author} {\bibfnamefont {J.~W.}\ \bibnamefont {Holt}}, \ and\ \bibinfo
  {author} {\bibfnamefont {C.~M.}\ \bibnamefont {Ko}},\ }\href {\doibase
  10.1016/j.physletb.2017.12.012} {\bibfield  {journal} {\bibinfo  {journal}
  {Phys. Lett. B}\ }\textbf {\bibinfo {volume} {777}},\ \bibinfo {pages} {73}
  (\bibinfo {year} {2018})},\ \Eprint {http://arxiv.org/abs/1703.00866}
  {arXiv:1703.00866 [nucl-th]} \BibitemShut {NoStop}%
\bibitem [{\citenamefont {Wang}\ \emph {et~al.}(2014)\citenamefont {Wang},
  \citenamefont {Guo}, \citenamefont {Li}, \citenamefont {Zhang}, \citenamefont
  {Leifels},\ and\ \citenamefont {Trautmann}}]{Wang:2014rva}%
  \BibitemOpen
  \bibfield  {author} {\bibinfo {author} {\bibfnamefont {Y.}~\bibnamefont
  {Wang}}, \bibinfo {author} {\bibfnamefont {C.}~\bibnamefont {Guo}}, \bibinfo
  {author} {\bibfnamefont {Q.}~\bibnamefont {Li}}, \bibinfo {author}
  {\bibfnamefont {H.}~\bibnamefont {Zhang}}, \bibinfo {author} {\bibfnamefont
  {Y.}~\bibnamefont {Leifels}}, \ and\ \bibinfo {author} {\bibfnamefont
  {W.}~\bibnamefont {Trautmann}},\ }\href {\doibase 10.1103/PhysRevC.89.044603}
  {\bibfield  {journal} {\bibinfo  {journal} {Phys. Rev. C}\ }\textbf {\bibinfo
  {volume} {89}},\ \bibinfo {pages} {044603} (\bibinfo {year} {2014})},\
  \Eprint {http://arxiv.org/abs/1403.7041} {arXiv:1403.7041 [nucl-th]}
  \BibitemShut {NoStop}%
\bibitem [{\citenamefont {Louren\c{c}o}\ \emph {et~al.}(2020)\citenamefont
  {Louren\c{c}o}, \citenamefont {Bhuyan}, \citenamefont {Lenzi}, \citenamefont
  {Dutra}, \citenamefont {Gonzalez-Boquera}, \citenamefont {Centelles},\ and\
  \citenamefont {Vi\~nas}}]{Lourenco:2020qft}%
  \BibitemOpen
  \bibfield  {author} {\bibinfo {author} {\bibfnamefont {O.}~\bibnamefont
  {Louren\c{c}o}}, \bibinfo {author} {\bibfnamefont {M.}~\bibnamefont
  {Bhuyan}}, \bibinfo {author} {\bibfnamefont {C.~H.}\ \bibnamefont {Lenzi}},
  \bibinfo {author} {\bibfnamefont {M.}~\bibnamefont {Dutra}}, \bibinfo
  {author} {\bibfnamefont {C.}~\bibnamefont {Gonzalez-Boquera}}, \bibinfo
  {author} {\bibfnamefont {M.}~\bibnamefont {Centelles}}, \ and\ \bibinfo
  {author} {\bibfnamefont {X.}~\bibnamefont {Vi\~nas}},\ }\href {\doibase
  10.1016/j.physletb.2020.135306} {\bibfield  {journal} {\bibinfo  {journal}
  {Phys. Lett. B}\ }\textbf {\bibinfo {volume} {803}},\ \bibinfo {pages}
  {135306} (\bibinfo {year} {2020})},\ \Eprint
  {http://arxiv.org/abs/2002.06242} {arXiv:2002.06242 [nucl-th]} \BibitemShut
  {NoStop}%
\bibitem [{\citenamefont {Angeli}\ and\ \citenamefont
  {Marinova}(2013)}]{ADNDT.99.69}%
  \BibitemOpen
  \bibfield  {author} {\bibinfo {author} {\bibfnamefont {I.}~\bibnamefont
  {Angeli}}\ and\ \bibinfo {author} {\bibfnamefont {K.}~\bibnamefont
  {Marinova}},\ }\href {\doibase https://doi.org/10.1016/j.adt.2011.12.006}
  {\bibfield  {journal} {\bibinfo  {journal} {At. Data Nucl. Data Tables}\
  }\textbf {\bibinfo {volume} {99}},\ \bibinfo {pages} {69} (\bibinfo {year}
  {2013})}\BibitemShut {NoStop}%
\bibitem [{\citenamefont {Goriely}\ \emph {et~al.}(2009)\citenamefont
  {Goriely}, \citenamefont {Hilaire}, \citenamefont {Girod},\ and\
  \citenamefont {Peru}}]{Goriely:2009zz}%
  \BibitemOpen
  \bibfield  {author} {\bibinfo {author} {\bibfnamefont {S.}~\bibnamefont
  {Goriely}}, \bibinfo {author} {\bibfnamefont {S.}~\bibnamefont {Hilaire}},
  \bibinfo {author} {\bibfnamefont {M.}~\bibnamefont {Girod}}, \ and\ \bibinfo
  {author} {\bibfnamefont {S.}~\bibnamefont {Peru}},\ }\href {\doibase
  10.1103/PhysRevLett.102.242501} {\bibfield  {journal} {\bibinfo  {journal}
  {Phys. Rev. Lett.}\ }\textbf {\bibinfo {volume} {102}},\ \bibinfo {pages}
  {242501} (\bibinfo {year} {2009})}\BibitemShut {NoStop}%
\bibitem [{\citenamefont {Robledo}\ \emph {et~al.}(2019)\citenamefont
  {Robledo}, \citenamefont {Rodr\'\i{}guez},\ and\ \citenamefont
  {Rodr\'\i{}guez-Guzm\'an}}]{Robledo:2018cdj}%
  \BibitemOpen
  \bibfield  {author} {\bibinfo {author} {\bibfnamefont {L.~M.}\ \bibnamefont
  {Robledo}}, \bibinfo {author} {\bibfnamefont {T.~R.}\ \bibnamefont
  {Rodr\'\i{}guez}}, \ and\ \bibinfo {author} {\bibfnamefont {R.~R.}\
  \bibnamefont {Rodr\'\i{}guez-Guzm\'an}},\ }\href {\doibase
  10.1088/1361-6471/aadebd} {\bibfield  {journal} {\bibinfo  {journal} {J.
  Phys. G}\ }\textbf {\bibinfo {volume} {46}},\ \bibinfo {pages} {013001}
  (\bibinfo {year} {2019})},\ \Eprint {http://arxiv.org/abs/1807.02518}
  {arXiv:1807.02518 [nucl-th]} \BibitemShut {NoStop}%
\bibitem [{\citenamefont {Sumi}\ \emph
  {et~al.}(2012{\natexlab{a}})\citenamefont {Sumi}, \citenamefont {Minomo},
  \citenamefont {Tagami}, \citenamefont {Kimura}, \citenamefont {Matsumoto},
  \citenamefont {Ogata}, \citenamefont {Shimizu},\ and\ \citenamefont
  {Yahiro}}]{PRC.85.064613}%
  \BibitemOpen
  \bibfield  {author} {\bibinfo {author} {\bibfnamefont {T.}~\bibnamefont
  {Sumi}}, \bibinfo {author} {\bibfnamefont {K.}~\bibnamefont {Minomo}},
  \bibinfo {author} {\bibfnamefont {S.}~\bibnamefont {Tagami}}, \bibinfo
  {author} {\bibfnamefont {M.}~\bibnamefont {Kimura}}, \bibinfo {author}
  {\bibfnamefont {T.}~\bibnamefont {Matsumoto}}, \bibinfo {author}
  {\bibfnamefont {K.}~\bibnamefont {Ogata}}, \bibinfo {author} {\bibfnamefont
  {Y.~R.}\ \bibnamefont {Shimizu}}, \ and\ \bibinfo {author} {\bibfnamefont
  {M.}~\bibnamefont {Yahiro}},\ }\href {\doibase 10.1103/PhysRevC.85.064613}
  {\bibfield  {journal} {\bibinfo  {journal} {Phys. Rev. C}\ }\textbf {\bibinfo
  {volume} {85}},\ \bibinfo {pages} {064613} (\bibinfo {year}
  {2012}{\natexlab{a}})}\BibitemShut {NoStop}%
\bibitem [{\citenamefont {Minomo}\ \emph {et~al.}(2010)\citenamefont {Minomo},
  \citenamefont {Ogata}, \citenamefont {Kohno}, \citenamefont {Shimizu},\ and\
  \citenamefont {Yahiro}}]{Minomo:2009ds}%
  \BibitemOpen
  \bibfield  {author} {\bibinfo {author} {\bibfnamefont {K.}~\bibnamefont
  {Minomo}}, \bibinfo {author} {\bibfnamefont {K.}~\bibnamefont {Ogata}},
  \bibinfo {author} {\bibfnamefont {M.}~\bibnamefont {Kohno}}, \bibinfo
  {author} {\bibfnamefont {Y.~R.}\ \bibnamefont {Shimizu}}, \ and\ \bibinfo
  {author} {\bibfnamefont {M.}~\bibnamefont {Yahiro}},\ }\href {\doibase
  10.1088/0954-3899/37/8/085011} {\bibfield  {journal} {\bibinfo  {journal} {J.
  Phys. G}\ }\textbf {\bibinfo {volume} {37}},\ \bibinfo {pages} {085011}
  (\bibinfo {year} {2010})},\ \Eprint {http://arxiv.org/abs/0911.1184}
  {arXiv:0911.1184 [nucl-th]} \BibitemShut {NoStop}%
\bibitem [{\citenamefont {Watanabe}\ \emph {et~al.}(2014)\citenamefont
  {Watanabe} \emph {et~al.}}]{Watanabe:2014zea}%
  \BibitemOpen
  \bibfield  {author} {\bibinfo {author} {\bibfnamefont {S.}~\bibnamefont
  {Watanabe}} \emph {et~al.},\ }\href {\doibase 10.1103/PhysRevC.89.044610}
  {\bibfield  {journal} {\bibinfo  {journal} {Phys. Rev. C}\ }\textbf {\bibinfo
  {volume} {89}},\ \bibinfo {pages} {044610} (\bibinfo {year} {2014})},\
  \Eprint {http://arxiv.org/abs/1404.2373} {arXiv:1404.2373 [nucl-th]}
  \BibitemShut {NoStop}%
\bibitem [{\citenamefont {Brieva}\ and\ \citenamefont
  {Rook}(1977{\natexlab{a}})}]{Brieva-Rook-1}%
  \BibitemOpen
  \bibfield  {author} {\bibinfo {author} {\bibfnamefont {F.~A.}\ \bibnamefont
  {Brieva}}\ and\ \bibinfo {author} {\bibfnamefont {J.~R.}\ \bibnamefont
  {Rook}},\ }\href@noop {} {\bibfield  {journal} {\bibinfo  {journal} {Nucl.
  Phys.}\ }\textbf {\bibinfo {volume} {291}},\ \bibinfo {pages} {299} (\bibinfo
  {year} {1977}{\natexlab{a}})}\BibitemShut {NoStop}%
\bibitem [{\citenamefont {Brieva}\ and\ \citenamefont
  {Rook}(1977{\natexlab{b}})}]{Brieva-Rook-2}%
  \BibitemOpen
  \bibfield  {author} {\bibinfo {author} {\bibfnamefont {F.~A.}\ \bibnamefont
  {Brieva}}\ and\ \bibinfo {author} {\bibfnamefont {J.~R.}\ \bibnamefont
  {Rook}},\ }\href@noop {} {\bibfield  {journal} {\bibinfo  {journal} {Nucl.
  Phys.}\ }\textbf {\bibinfo {volume} {291}},\ \bibinfo {pages} {317} (\bibinfo
  {year} {1977}{\natexlab{b}})}\BibitemShut {NoStop}%
\bibitem [{\citenamefont {Brieva}\ and\ \citenamefont
  {Rook}(1978)}]{Brieva-Rook-3}%
  \BibitemOpen
  \bibfield  {author} {\bibinfo {author} {\bibfnamefont {F.~A.}\ \bibnamefont
  {Brieva}}\ and\ \bibinfo {author} {\bibfnamefont {J.~R.}\ \bibnamefont
  {Rook}},\ }\href@noop {} {\bibfield  {journal} {\bibinfo  {journal} {Nucl.
  Phys.}\ }\textbf {\bibinfo {volume} {297}},\ \bibinfo {pages} {206} (\bibinfo
  {year} {1978})}\BibitemShut {NoStop}%
\bibitem [{\citenamefont {de~Vries}\ \emph {et~al.}(1987)\citenamefont
  {de~Vries}, \citenamefont {de~Jager},\ and\ \citenamefont
  {de~Vries}}]{C12-density}%
  \BibitemOpen
  \bibfield  {author} {\bibinfo {author} {\bibfnamefont {H.}~\bibnamefont
  {de~Vries}}, \bibinfo {author} {\bibfnamefont {C.~W.}\ \bibnamefont
  {de~Jager}}, \ and\ \bibinfo {author} {\bibfnamefont {C.}~\bibnamefont
  {de~Vries}},\ }\href@noop {} {\bibfield  {journal} {\bibinfo  {journal} {At.
  Data Nucl. Data Tables}\ }\textbf {\bibinfo {volume} {36}},\ \bibinfo {pages}
  {495} (\bibinfo {year} {1987})}\BibitemShut {NoStop}%
\bibitem [{\citenamefont {Takechi}\ \emph {et~al.}(2021)\citenamefont
  {Takechi}, \citenamefont {Wakasa}, \citenamefont {Tagami}, \citenamefont
  {Matsui},\ and\ \citenamefont {Yahiro}}]{Takechi:2020snn}%
  \BibitemOpen
  \bibfield  {author} {\bibinfo {author} {\bibfnamefont {M.}~\bibnamefont
  {Takechi}}, \bibinfo {author} {\bibfnamefont {T.}~\bibnamefont {Wakasa}},
  \bibinfo {author} {\bibfnamefont {S.}~\bibnamefont {Tagami}}, \bibinfo
  {author} {\bibfnamefont {J.}~\bibnamefont {Matsui}}, \ and\ \bibinfo {author}
  {\bibfnamefont {M.}~\bibnamefont {Yahiro}},\ }\href {\doibase
  10.1016/j.rinp.2021.104923} {\bibfield  {journal} {\bibinfo  {journal}
  {Results Phys.}\ }\textbf {\bibinfo {volume} {31}},\ \bibinfo {pages}
  {104923} (\bibinfo {year} {2021})},\ \Eprint
  {http://arxiv.org/abs/2009.00796} {arXiv:2009.00796 [nucl-th]} \BibitemShut
  {NoStop}%
\bibitem [{\citenamefont {Sumi}\ \emph
  {et~al.}(2012{\natexlab{b}})\citenamefont {Sumi}, \citenamefont {Minomo},
  \citenamefont {Tagami}, \citenamefont {Kimura}, \citenamefont {Matsumoto},
  \citenamefont {Ogata}, \citenamefont {Shimizu},\ and\ \citenamefont
  {Yahiro}}]{Sumi:2012fr}%
  \BibitemOpen
  \bibfield  {author} {\bibinfo {author} {\bibfnamefont {T.}~\bibnamefont
  {Sumi}}, \bibinfo {author} {\bibfnamefont {K.}~\bibnamefont {Minomo}},
  \bibinfo {author} {\bibfnamefont {S.}~\bibnamefont {Tagami}}, \bibinfo
  {author} {\bibfnamefont {M.}~\bibnamefont {Kimura}}, \bibinfo {author}
  {\bibfnamefont {T.}~\bibnamefont {Matsumoto}}, \bibinfo {author}
  {\bibfnamefont {K.}~\bibnamefont {Ogata}}, \bibinfo {author} {\bibfnamefont
  {Y.~R.}\ \bibnamefont {Shimizu}}, \ and\ \bibinfo {author} {\bibfnamefont
  {M.}~\bibnamefont {Yahiro}},\ }\href {\doibase 10.1103/PhysRevC.85.064613}
  {\bibfield  {journal} {\bibinfo  {journal} {Phys. Rev. C}\ }\textbf {\bibinfo
  {volume} {85}},\ \bibinfo {pages} {064613} (\bibinfo {year}
  {2012}{\natexlab{b}})},\ \Eprint {http://arxiv.org/abs/1201.2497}
  {arXiv:1201.2497 [nucl-th]} \BibitemShut {NoStop}%
\bibitem [{HP:()}]{HP:NuDat2.8}%
  \BibitemOpen
  \href@noop {} {}\bibinfo {note}
  {{h}ttps://www.nndc.bnl.gov/nudat2/}\BibitemShut {NoStop}%
\end{thebibliography}%

\onecolumngrid

\squeezetable
\begin{table}[h]
\caption{Properties of 206 EoSs (1). The symbol $^\ddagger$ is our results, while 
$^\dagger$  denotes the results of self-consistent calculations.   
}
\label{table-I}
\begin{tabular}{c|ccccccc|cc}
\hline
 & m*/m & K &J & L & Ksym & Rskin-208 & Rskin-48 & Refs. \cr
\hline	      
APR, E0019& & 266.000  & 32.600  & 57.600  &  & 0.160   & 0.160$^\dagger$  
& \cite{Akmal:1998cf,Brown:2000pd, Ishizuka:2014jsa}   & \cr
BHF-1 &  & 195.500  & 34.300  & 66.550  & -31.300  & 0.200$^\dagger$  & 0.183$^\dagger$  & 
\cite{Ducoin:2010as}  & \cr
BSk14 & 0.800  & 239.380  & 30.000  & 43.910  & -152.030  & 0.164$^\dagger$  & 0.162$^\dagger$  & 
\cite{Ducoin:2010as,Dutra:2012mb}  &\cr 
BSk16 & 0.800  & 241.730  & 30.000  & 34.870  & -187.390  & 0.149$^\dagger$  & 0.154$^\dagger$  & 
\cite{Ducoin:2010as,Dutra:2012mb} & \cr
BSk17 & 0.800  & 241.740  & 30.000  & 36.280  & -181.860  & 0.151$^\dagger$  & 0.156$^\dagger$  & 
\cite{Ducoin:2010as,Dutra:2012mb}  & \cr
BSk20 & 0.800  & 241.400  & 30.000  & 37.400  & -136.500  & 0.153$^\dagger$  & 0.157$^\dagger$  &  \cite{Fortin:2016hny,Dutra:2012mb}  & \cr
BSk21 & 0.800  & 245.800  & 30.000  & 46.600  & -37.200  & 0.168$^\dagger$  & 0.165$^\dagger$  &  \cite{Fortin:2016hny,Dutra:2012mb}  & \cr
BSk22 &  & 245.900  & 32.000  & 68.500  & 13.000  & 0.204$^\dagger$  & 0.185$^\dagger$  &  
\cite{Fortin:2016hny}  & \cr
BSk23 &  & 245.700  & 31.000  & 57.800  & -11.300  & 0.186$^\dagger$  & 0.175$^\dagger$  & 
\cite{Fortin:2016hny,Ishizuka:2014jsa}   & \cr
BSk24 &  & 245.500  & 30.000  & 46.400  & -37.600  & 0.168$^\dagger$  & 0.165$^\dagger$  & 
\cite{Fortin:2016hny}   & \cr
BSk25 &  & 236.000  & 29.000  & 36.900  & -28.500  & 0.152$^\dagger$  & 0.156$^\dagger$   & 
\cite{Fortin:2016hny}   & \cr
BSk26 &  & 240.800  & 30.000  & 37.500  & -135.600  & 0.153$^\dagger$  & 0.157$^\dagger$  & 
\cite{Fortin:2016hny}   & \cr
BSR2 &  & 239.900  & 31.500  & 62.000  & -3.100  & 0.193$^\dagger$  & 0.179$^\dagger$  & 
\cite{Fortin:2016hny}   & \cr
BSR6 &  & 235.800  & 35.600  & 85.700  & -49.600  & 0.231$^\dagger$  & 0.200$^\dagger$  & 
\cite{Fortin:2016hny}   & \cr
D1 &  & 229.400  & 30.700  & 18.360  & -274.600  & 0.122$^\dagger$  & 0.139$^\dagger$  & 
\cite{Gonzalez-Boquera:2017uep}   & \cr
D1AS &  & 229.400  & 31.300  & 66.550  & -89.100  & 0.200$^\dagger$  & 0.183$^\dagger$  & 
\cite{Gonzalez-Boquera:2017uep}  & \cr
D1M & 0.746$^\ddagger$  & 224.958$^\ddagger$  & 28.552$^\ddagger$  & 24.966$^\ddagger$  & -133.692$^\ddagger$  & 0.113$^\ddagger$  & 0.147$^\ddagger$  & 
\cite{Gonzalez-Boquera:2017rzy}  & \cr
D1M* & 0.746$^\ddagger$  & 225.365$^\ddagger$  & 30.249$^\ddagger$  & 43.311$^\ddagger$  & -47.793$^\ddagger$  & 0.134$^\ddagger$  & 0.158$^\ddagger$  & 
\cite{Gonzalez-Boquera:2017rzy}   & \cr
D1MK & 0.746$^\ddagger$  & 225.400$^\ddagger$  & 33.000$^\ddagger$  & 55.000$^\ddagger$  & -37.275$^\ddagger$  & 0.158$^\ddagger$  & 0.171$^\ddagger$  & TW  & \cr
D1N & 0.748$^\ddagger$  & 225.525$^\ddagger$  & 29.594$^\ddagger$  & 33.665$^\ddagger$  & -168.750$^\ddagger$  & 0.144$^\ddagger$  & 0.171$^\ddagger$  & 
\cite{Gonzalez-Boquera:2017rzy}  & \cr
D1P & 0.672$^\ddagger$  & 250.860$^\ddagger$  & 32.418$^\ddagger$  & 49.827$^\ddagger$  & -157.419$^\ddagger$  & 0.179$^\ddagger$  & 0.157$^\ddagger$  & 
\cite{D1P-99,Gonzalez-Boquera:2017uep}  & \cr
D1PK & 0.700$^\ddagger$  & 260.000$^\ddagger$  & 33.000$^\ddagger$  & 55.000$^\ddagger$  & -150.000$^\ddagger$  & 0.182$^\ddagger$  & 0.181$^\ddagger$  & TW  & \cr
D1S & 0.697$^\ddagger$  & 202.856$^\ddagger$  & 31.125$^\ddagger$  & 22.558$^\ddagger$  & -241.797$^\ddagger$  & 0.137$^\ddagger$  & 0.159$^\ddagger$  & 
\cite{Gonzalez-Boquera:2017rzy,Inakura:2015cla}  & \cr
D2 & 0.738  & 209.300  & 31.130  & 44.850  &  & 0.165$^\dagger$  & 0.163$^\dagger$  & 
\cite{Gonzalez-Boquera:2017rzy}  & \cr
D250 &  & 249.900  & 31.570  & 24.820  & -289.400  & 0.133$^\dagger$  & 0.145$^\dagger$  & 
\cite{Gonzalez-Boquera:2017uep}   & \cr
D260 &  & 259.500  & 30.110  & 17.570  & -298.700  & 0.121$^\dagger$  & 0.139$^\dagger$  & 
\cite{Gonzalez-Boquera:2017uep}   & \cr
D280 &  & 285.200  & 33.140  & 46.530  & -211.900  & 0.168$^\dagger$  & 0.165$^\dagger$  & 
\cite{Gonzalez-Boquera:2017uep}   & \cr
D300 &  & 299.100  & 31.220  & 25.840  & -315.100  & 0.135$^\dagger$  & 0.146$^\dagger$  & 
\cite{Gonzalez-Boquera:2017uep}   & \cr
DD &  & 241.000  & 31.700  & 56.000  & -95.000  & 0.183$^\dagger$  & 0.173$^\dagger$  &  
\cite{Chen:2010qx}  & \cr
DD-F &  & 223.000  & 31.600  & 56.000  & -140.000  & 0.183$^\dagger$  & 0.173$^\dagger$  & 
\cite{Chen:2010qx}  & \cr
DD-ME1 &  & 245.000  & 33.100  & 55.000  & -101.000  & 0.203$^\dagger$  & 0.193$^\dagger$  & 
\cite{Chen:2010qx,Zhao:2016ujh,Ducoin:2010as,Ishizuka:2014jsa}  & \cr
DD-ME2 &  & 251.000  & 32.300  & 51.240  & -87.000  & 0.203$^\dagger$  & 0.187$^\dagger$  & 
\cite{Chen:2010qx,Zhao:2016ujh,Ducoin:2010as,Fortin:2016hny}  & \cr
DD-PC1 &  &  &  & 67.799  &  & 0.203$^\dagger$  & 0.195$^\dagger$  & 
\cite{Zhao:2016ujh,Ducoin:2010as}  & \cr
Ducoin &  & 240.200  & 32.760  & 55.300  & -124.700  & 0.182$^\dagger$  & 0.173$^\dagger$  & 
\cite{Ducoin:2010as}   & \cr
E0008(TMA) &  & 318.000  & 30.660  & 90.140  &  & 0.239$^\dagger$  & 0.204$^\dagger$  & 
\cite{Ishizuka:2014jsa}  & \cr
E0009 &  & 280.000  & 32.500  & 88.700  &  & 0.236$^\dagger$  & 0.203$^\dagger$  & 
\cite{Ishizuka:2014jsa,Oertel:2016bki}  & \cr
E0015 &  & 216.700  & 30.030  & 45.780  &  & 0.167  & 0.164  & 
\cite{Ishizuka:2014jsa}  & \cr
E0024 &  & 244.500  & 33.100  & 55.000  &  & 0.182$^\dagger$  & 0.172  & 
\cite{Ishizuka:2014jsa}  & \cr
E0025 &  & 211.000  & 31.600  & 107.400  &  & 0.267$^\dagger$  & 0.220$^\dagger$  & 
\cite{Ishizuka:2014jsa}  & \cr
E0036 &  & 281.000  & 36.900  & 110.800  &  & 0.272$^\dagger$  & 0.223$^\dagger$  & 
\cite{Ishizuka:2014jsa}  & \cr
es25 &  & 211.730  & 25.000  & 27.749$^\dagger$  &  & 0.138  & 0.148$^\dagger$  & 
\cite{Steiner:2004fi}   & \cr
es275 &  & 205.330  & 27.500  & 48.549$^\dagger$  &  & 0.171  & 0.167$^\dagger$  & 
\cite{Steiner:2004fi,Fortin:2016hny}  & \cr
es30 &  & 215.360  & 30.000  & 69.603$^\dagger$  &  & 0.205  & 0.186$^\dagger$  & 
\cite{Steiner:2004fi}  & \cr
es325 &  & 212.450  & 32.500  & 81.925$^\dagger$  &  & 0.225  & 0.197$^\dagger$  & 
\cite{Steiner:2004fi}  & \cr
es35 &  & 209.970  & 34.937  & 96.182$^\dagger$  &  & 0.248  & 0.210$^\dagger$  & 
\cite{Steiner:2004fi}  & \cr
FKVW &  & 379.000  & 33.100  & 80.000  & 11.000  & 0.222$^\dagger$  & 0.195$^\dagger$  & 
\cite{Chen:2010qx}   & \cr
FSU &  & 230.000  & 32.590  & 60.500  & -51.300  & 0.210  & 0.188$^\dagger$  & 
\cite{Fattoyev:2013yaa,Ducoin:2010as}  & \cr
FSUgold &  & 229.000  & 32.500  & 60.000  & -52.000  & 0.210  & 0.200  & 
\cite{Chen:2010qx,Oertel:2016bki,Piekarewicz:2007dx}    & \cr
FSUgold2.1 &  & 230.000  & 32.590  & 60.500  &  & 0.191$^\dagger$  & 0.177$^\dagger$  & 
\cite{Ishizuka:2014jsa,Oertel:2016bki}   & \cr
GM1 &  & 299.700  & 32.480  & 93.870  & 17.890  & 0.245$^\dagger$  & 0.207$^\dagger$  & 
\cite{Ducoin:2010as,Fortin:2016hny}   & \cr
GM3 &  & 239.900  & 32.480  & 89.660  & -6.470  & 0.238$^\dagger$  & 0.204$^\dagger$  & 
\cite{Ducoin:2010as}  & \cr
Gs &  & 237.570  & 31.384  & 89.304$^\dagger$  &  & 0.237  & 0.203$^\dagger$  & 
\cite{Steiner:2004fi}  & \cr
GSkI &  & 230.210  & 32.030  & 63.450  & -95.290  & 0.195$^\dagger$  & 0.180$^\dagger$  & 
\cite{Dutra:2012mb}   & \cr
GSkII & 0.790  & 233.400  & 30.490  & 48.630  & -157.830  & 0.171$^\dagger$  & 0.167$^\dagger$  & \cite{Dutra:2012mb}   & \cr
GT2 &  & 228.100  & 33.940  & 5.020  & -445.900  & 0.101$^\dagger$  & 0.127$^\dagger$  & 
\cite{Gonzalez-Boquera:2017uep}  & \cr
G$\sigma$ &  & 237.290  & 31.370  & 94.020  & 13.990  & 0.245$^\dagger$  & 0.208$^\dagger$  & 
\cite{Ducoin:2010as}  & \cr
HA &  & 233.000  & 30.700  & 55.000  & -135.000  & 0.182$^\dagger$  & 0.172$^\dagger$  & 
\cite{Chen:2010qx}  & \cr
HFB-17 &  &  &  & 36.300  &  & 0.151  & 0.155$^\dagger$  & 
\cite{RocaMaza:2011pm}   & \cr
HFB-8 &  &  &  & 14.800  &  & 0.115  & 0.135$^\dagger$  & 
\cite{RocaMaza:2011pm}  & \cr
HS(DD2) &  & 243.000  & 31.700  & 55.000  & -93.200  & 0.182$^\dagger$  & 0.172$^\dagger$  & 
\cite{Oertel:2016bki,Fortin:2016hny}  & \cr
IU-FSU &  & 231.200  & 31.300  & 47.200  & 28.700  & 0.160  & 0.160$^\dagger$  & 
\cite{Fattoyev:2013yaa, Oertel:2016bki}    & \cr
KDE0v1 & 0.740  & 227.540  & 34.580  & 54.690  & -127.120  & 0.181$^\dagger$  & 0.172$^\dagger$  & 
\cite{Dutra:2012mb}   & \cr
KDE0v1-B & 0.790  & 216.000  & 34.900  & 61.000  &  & 0.192  & 0.172  & 
\cite{Brown:2013pwa}  & \cr
KDE0v1-T & 0.810  & 217.000  & 34.600  & 72.000  & -40.000  & 0.200  & 0.178  & 
\cite{Tsang:2019ymt}   & \cr
\hline
\end{tabular}
\end{table} 
\setcounter{table}{0}
\begin{table}[h]
\caption{Properties of 206 EoSs (2).
}
\begin{tabular}{c|ccccccc|cc}
\hline
 & m*/m & K &J & L & Ksym & Rskin-208 & Rskin-48 & Refs. \cr
\hline
LNS & 0.830  & 210.780  & 33.430  & 61.450  & -127.360  & 0.192$^\dagger$  & 0.178$^\dagger$  & 
\cite{Dutra:2012mb,Ducoin:2010as}     & \cr
LS180 &  & 180.000  & 28.600  & 73.800  &  & 0.212$^\dagger$  & 0.189$^\dagger$  & 
\cite{Dutra:2012mb,Ishizuka:2014jsa,Oertel:2016bki}   & \cr
LS220 &  & 220.000  & 28.600  & 73.800  &  & 0.212$^\dagger$  & 0.189$^\dagger$  & 
\cite{Dutra:2012mb,Ishizuka:2014jsa,Oertel:2016bki}   & \cr
LS375 &  & 375.000  & 28.600  & 73.800  &  & 0.212$^\dagger$  & 0.189$^\dagger$  & 
\cite{Dutra:2012mb,Ishizuka:2014jsa,Oertel:2016bki}  & \cr
Ly5 &  & 229.940  & 32.010  & 45.243$^\dagger$  &  & 0.166  & 0.164$^\dagger$  & 
\cite{Steiner:2004fi}   & \cr
M3Y-P6 &  & 239.700  & 32.100  & 44.600  & -165.300  & 0.165$^\dagger$  & 0.163$^\dagger$  & 
\cite{Inakura:2015cla,Lim:2013tqa}  & \cr
M3Y-P7 &  & 254.700  & 31.700  & 51.500  & -127.800  & 0.176$^\dagger$  & 0.169$^\dagger$  & 
\cite{Inakura:2015cla,Fortin:2016hny}  & \cr
MSk3 & 1.000  & 233.250  & 28.000  & 7.040  & -283.520  & 0.111  & 0.128  &  
\cite{Dutra:2012mb,Zhao:2016ujh}   & \cr
MSk6 & 1.050  & 231.170  & 28.000  & 9.630  & -274.330  & 0.118  & 0.130  &  
\cite{Dutra:2012mb,Zhao:2016ujh}   & \cr
MSk7 & 1.050  & 385.360  & 27.950  & 9.400  & -274.630  & 0.116  & 0.136$^\dagger$  & 
\cite{RocaMaza:2011pm,Dutra:2012mb}   & \cr
MSL0 & 0.800  & 230.000  & 30.000  & 60.000  & -99.330  & 0.180  & 0.171$^\dagger$  & 
\cite{Chen:2010qx,Dutra:2012mb,Wang:2014rva}  & \cr
NL1 &  & 212.000  & 43.500  & 140.000  & 143.000  & 0.319  & 0.247  & 
\cite{Chen:2010qx,Zhao:2016ujh}   & \cr
NL2 &  & 401.000  & 44.000  & 130.000  & 20.000  & 0.304  & 0.243  & 
\cite{Chen:2010qx,Zhao:2016ujh}   & \cr
NL3 &  & 271.000  & 37.300  & 118.000  & 100.000  & 0.280  & 0.230  & 
\cite{Chen:2010qx,Fattoyev:2013yaa,Centelles:2010qh,Piekarewicz:2007dx,Ducoin:2010as,Fortin:2016hny} 
 & \cr
NL3* &  &  &  & 119.769$^\dagger$  &  & 0.287  & 0.230  & 
\cite{Zhao:2016ujh}  & \cr
NL3$\omega \rho$ &  & 271.600  & 31.700  & 55.500  & -7.600  & 0.183$^\dagger$  & 0.173$^\dagger$  & \cite{Fortin:2016hny}  & \cr
NL4 &  & 270.350  & 36.239  & 111.649$^\dagger$  &  & 0.273  & 0.223$^\dagger$  & 
\cite{Steiner:2004fi}   & \cr
NL-SH &  & 356.000  & 36.100  & 114.000  & 80.000  & 0.263  & 0.214  & 
\cite{Chen:2010qx,Zhao:2016ujh}   & \cr
NL$\rho$ &  & 240.000  & 30.300  & 85.000  & 3.000  & 0.230$^\dagger$  & 0.199$^\dagger$  & 
\cite{Chen:2010qx}  & \cr
NL$\omega \rho$(025) &  & 270.700  & 32.350  & 61.050  & -34.360  & 0.192$^\dagger$  & 0.178$^\dagger$  & \cite{Ducoin:2010as}  & \cr
NRAPR & 0.690  & 225.700  & 32.787  & 59.630  & -123.320  & 0.190  & 0.177$^\dagger$  & 
\cite{Steiner:2004fi,Dutra:2012mb}   & \cr
NRAPR-B & 0.850  & 225.000  & 35.100  & 61.000  &  & 0.193  & 0.178  & 
\cite{Brown:2013pwa}  & \cr
NRAPR-T & 0.730  & 221.000  & 34.100  & 70.000  & -46.000  & 0.195  & 0.181  & 
\cite{Tsang:2019ymt}  & \cr
PC-F1 &  & 255.000  & 37.800  & 117.000  & 75.000  & 0.269  & 0.225  & 
\cite{Chen:2010qx,Zhao:2016ujh}  & \cr
PC-F2 &  & 256.000  & 37.600  & 116.000  & 65.000  & 0.281$^\dagger$  & 0.227$^\dagger$  & 
\cite{Chen:2010qx,Zhao:2016ujh,Wang:2014rva}  & \cr
PC-F3 &  & 256.000  & 38.300  & 119.000  & 74.000  & 0.285$^\dagger$  & 0.230$^\dagger$  & 
\cite{Chen:2010qx,Zhao:2016ujh} & \cr
PC-F4 &  & 255.000  & 37.700  & 119.000  & 98.000  & 0.285$^\dagger$  & 0.230$^\dagger$  & 
\cite{Chen:2010qx}  & \cr
PC-LA &  & 263.000  & 37.200  & 108.000  & -61.000  & 0.268$^\dagger$  & 0.220$^\dagger$  & 
\cite{Chen:2010qx}  & \cr
PC-PK1 &  &  &  & 101.478  &  & 0.257  & 0.220  & 
\cite{Zhao:2016ujh}  & \cr
PK1 &  & 282.000  & 37.600  & 116.000  & 55.000  & 0.277  & 0.223  & 
\cite{Chen:2010qx,Zhao:2016ujh}   & \cr
PKDD &  & 263.000  & 36.900  & 90.000  & -80.000  & 0.253  & 0.214  & 
\cite{Chen:2010qx,Zhao:2016ujh}   & \cr
RAPR &  & 276.700  & 33.987  & 66.958$^\dagger$  &  & 0.201  & 0.183$^\dagger$  & 
\cite{Steiner:2004fi}   & \cr
RATP &  & 239.580  & 29.260  & 32.390  & -191.250  & 0.145$^\dagger$  & 0.152$^\dagger$  & 
\cite{Ducoin:2010as}  & \cr
rDD-ME2 &  &  &  & 51.300  &  & 0.193  & 0.179$^\dagger$  & 
\cite{RocaMaza:2011pm}   & \cr
rFSUGold &  &  &  & 60.500  &  & 0.207  & 0.186$^\dagger$  & 
\cite{RocaMaza:2011pm,Fortin:2016hny}  & \cr
rG2 &  &  &  & 100.700  &  & 0.257  & 0.214$^\dagger$  & 
\cite{RocaMaza:2011pm}  & \cr
rNL1 &  &  &  & 140.100  &  & 0.321  & 0.250$^\dagger$  & 
\cite{RocaMaza:2011pm}  & \cr
rNL3 &  &  &  & 118.500  &  & 0.280  & 0.227$^\dagger$  & 
\cite{RocaMaza:2011pm}   & \cr
rNL3* &  &  &  & 122.600  &  & 0.288  & 0.232$^\dagger$  & \cite{RocaMaza:2011pm}  & \cr
rNLC &  &  &  & 108.000  &  & 0.263  & 0.218$^\dagger$  & 
\cite{RocaMaza:2011pm}  & \cr
rNL-RA1 &  &  &  & 115.400  &  & 0.274  & 0.224$^\dagger$  & 
\cite{RocaMaza:2011pm}  & \cr
rNL-SH &  &  &  & 113.600  &  & 0.266  & 0.219$^\dagger$  & 
\cite{RocaMaza:2011pm}  & \cr
rNL-Z &  &  &  & 133.300  &  & 0.307  & 0.242$^\dagger$  & 
\cite{RocaMaza:2011pm}  & \cr
Rs &  & 237.660  & 30.593  & 80.096$^\dagger$  & -9.100  & 0.222  & 0.195$^\dagger$  & 
\cite{Steiner:2004fi,Fortin:2016hny}  & \cr
rTM1 &  &  &  & 110.800  &  & 0.271  & 0.222$^\dagger$  & 
\cite{RocaMaza:2011pm}  & \cr
R$\sigma$&  & 237.410  & 30.580  & 85.700  & -9.130  & 0.231$^\dagger$  & 0.200$^\dagger$  & 
\cite{Ducoin:2010as}  & \cr
S271 &  & 271.000  & 35.927  & 97.541$^\dagger$  &  & 0.251  & 0.211$^\dagger$  & 
\cite{Steiner:2004fi}   & \cr
SFHo &  & 245.000  & 31.600  & 47.100  &  & 0.169$^\dagger $ & 0.165$^\dagger$  & 
\cite{Oertel:2016bki}  & \cr
SFHx &  & 239.000  & 28.700  & 23.200  &  & 0.130$^\dagger$  & 0.144$^\dagger$  & 
\cite{Oertel:2016bki}  & \cr
SGI & 0.610  & 262.000  & 28.300  & 63.900  & -51.990  & 0.196$^\dagger$  & 0.180$^\dagger$  & 
\cite{Steiner:2004fi,Lim:2013tqa,Dutra:2012mb}  & \cr
SGII & 0.790  & 214.700  & 26.830  & 37.620  & -145.920  & 0.136  & 0.147$^\dagger$  & 
\cite{Ducoin:2010as,RocaMaza:2011pm,Inakura:2015cla,Dutra:2012mb} & \cr
SII & 0.580  & 341.400  & 34.160  & 50.020  & -265.720  & 0.196  & 0.177  &   
\cite{Zhao:2016ujh}  & \cr
SIII & 0.760  & 355.37 & 28.160  & 9.910  & -393.730  & 0.137  & 0.125  & 
\cite{Dutra:2012mb,Zhao:2016ujh}   & \cr
Sk$\chi$m &  & 230.400  & 30.940  & 45.600  &  & 0.167  & 0.164$^\dagger$  & 
\cite{Zhang:2017hvh,Dutra:2012mb}   & \cr
SK255 &  & 254.960  & 37.400  & 95.000  & -58.300  & 0.247$^\dagger$  & 0.208$^\dagger$  & 
\cite{Fortin:2016hny}  & \cr
SK272 &  & 271.550  & 37.400  & 91.700  & -67.800  & 0.241$^\dagger$  & 0.205$^\dagger$  & 
\cite{Fortin:2016hny}  & \cr
Ska & 0.610  & 263.160  & 32.910  & 74.620  & -78.460  & 0.214$^\dagger$  & 0.190$^\dagger$  & 
\cite{RocaMaza:2011pm,Dutra:2012mb,Fortin:2016hny,Oertel:2016bki,Inakura:2015cla,Wang:2014rva}  & \cr
Ska25-B & 0.990  & 219.000  & 32.500  & 51.000  &  & 0.176  & 0.170  & 
\cite{Brown:2013pwa}  & \cr
Ska25s20 & 0.980  & 220.750  & 33.780  & 63.810  & -118.220  & 0.196$^\dagger$  & 0.180$^\dagger$  & \cite{Dutra:2012mb}  & \cr
Ska25-T & 0.980  & 220.000  & 31.900  & 59.000  & -59.000  & 0.183  & 0.176  & 
\cite{Tsang:2019ymt}  & \cr
Ska35-B & 1.000  & 244.000  & 32.800  & 54.000  &  & 0.180  & 0.172  & 
\cite{Brown:2013pwa}  & \cr
Ska35s20 & 1.000  & 240.270  & 33.570  & 64.830  & -120.320  & 0.198$^\dagger$  & 0.181$^\dagger$  & \cite{Dutra:2012mb,Wang:2014rva}  & \cr
Ska35-T & 0.990  & 238.000  & 32.000  & 58.000  & -84.000  & 0.184  & 0.177  & 
\cite{Tsang:2019ymt,Dutra:2012mb,Ducoin:2010as}   & \cr
\hline
\end{tabular}
\end{table} 

\setcounter{table}{0}
\begin{table}[h]
\caption{Properties of 206 EoSs (3).
}
\begin{tabular}{c|ccccccc|cc}
\hline
 & m*/m & K &J & L & Ksym & Rskin-208 & Rskin-48 & Refs. \cr
\hline
SKb & 0.610  & 263.000  & 33.880  & 47.600  & -78.500  & 0.170$^\dagger$  & 0.166$^\dagger$  & 
\cite{Fortin:2016hny,Dutra:2012mb}  & \cr
SkI1 & 0.690  & 242.750  & 37.530  & 161.050  & 234.670  & 0.353$^\dagger$  & 0.268$^\dagger$  & 
\cite{Wang:2014rva,Dutra:2012mb}  & \cr
SkI2 & 0.680  & 240.700  & 33.400  & 104.300  & 70.600  & 0.262$^\dagger$  & 0.217$^\dagger$  &  \cite{Ducoin:2010as,Fortin:2016hny,Dutra:2012mb,Inakura:2015cla}   & \cr
SkI3 & 0.580  & 258.000  & 34.800  & 100.500  & 72.900  & 0.255$^\dagger$  & 0.213$^\dagger$  & 
\cite{Ducoin:2010as,Fortin:2016hny,Dutra:2012mb,Inakura:2015cla} & \cr
SkI4 & 0.650  & 247.700  & 29.500  & 60.400  & -40.600  & 0.191$^\dagger$  & 0.177$^\dagger$  & 
\cite{Ducoin:2010as,Fortin:2016hny,Dutra:2012mb,Inakura:2015cla,Lim:2013tqa}  & \cr
SkI5 & 0.580  & 255.800  & 36.697  & 129.300  & 159.500  & 0.272  & 0.214  & 
\cite{Steiner:2004fi,Ducoin:2010as,Fortin:2016hny,Dutra:2012mb,Inakura:2015cla}   & \cr
SkI6 & 0.640  & 248.650  & 30.090  & 59.700  & -47.270  & 0.189$^\dagger$  & 0.177$^\dagger$  & 
\cite{Ducoin:2010as,Fortin:2016hny,Dutra:2012mb}   & \cr
SkM* & 0.790  & 216.610  & 30.030  & 45.780  & -155.940  & 0.170  & 0.155  & 
\cite{RocaMaza:2011pm,Dutra:2012mb,Inakura:2015cla,Zhao:2016ujh}  & \cr
SkM*-B & 0.780  & 218.000  & 34.200  & 58.000  &  & 0.187  & 0.175  & 
\cite{Brown:2013pwa}  & \cr
SkM*-T & 0.790  & 219.000  & 33.700  & 65.000  & -65.000  & 0.187  & 0.179  & 
\cite{Tsang:2019ymt,} & \cr
SkMP & 0.650  & 230.930  & 29.890  & 70.310  & -49.820  & 0.197  & 0.167  & 
\cite{RocaMaza:2011pm,Ducoin:2010as,Steiner:2004f,Fortin:2016hny}  & \cr
SkO & 0.900  & 223.390  & 31.970  & 79.140  & -43.170  & 0.221$^\dagger$  & 0.194$^\dagger$  & 
\cite{Ducoin:2010as,Dutra:2012mb}   & \cr
SKOp & 0.900  & 222.360  & 31.950  & 68.940  & -78.820  & 0.204$^\dagger$  & 0.185$^\dagger$ & \cite{Dutra:2012mb,Fortin:2016hny}  & \cr
SKP & 1.000  & 200.970  & 30.000  & 19.680  & -266.600  & 0.144  & 0.144  &  
\cite{Dutra:2012mb,Zhao:2016ujh}   & \cr
SKRA & 0.750  & 216.980  & 31.320  & 53.040  & -139.280  & 0.179$^\dagger$  & 0.171$^\dagger$  & 
\cite{Dutra:2012mb}   & \cr
SKRA-B & 0.790  & 212.000  & 33.700  & 55.000  &  & 0.181  & 0.172  & 
\cite{Brown:2013pwa,Dutra:2012mb}  & \cr
SKRA-T & 0.800  & 213.000  & 33.400  & 65.000  & -55.000  & 0.190  & 0.179  & 
\cite{Tsang:2019ymt,Dutra:2012mb}   & \cr
Sk-Rs &  &  &  & 85.700  &  & 0.215  & 0.191$^\dagger$  & 
\cite{RocaMaza:2011pm}  & \cr
SkSM* &  &  &  & 65.500  &  & 0.197  & 0.181$^\dagger$  & 
\cite{RocaMaza:2011pm}  & \cr
SkT1 & 1.000  & 236.160  & 32.020  & 56.180  & -134.830  & 0.184$^\dagger$  & 0.173$^\dagger$  &   
\cite{Dutra:2012mb}   & \cr
SkT1-B & 0.970  & 242.000  & 33.300  & 56.000  &  & 0.183  & 0.172  & 
\cite{Brown:2013pwa}  & \cr
SkT1-T & 0.970  & 238.000  & 32.600  & 63.000  & -70.000  & 0.190  & 0.179  & 
\cite{Tsang:2019ymt,Dutra:2012mb}   & \cr
SkT2 & 1.000  & 235.730  & 32.000  & 56.160  & -134.670  & 0.184$^\dagger$  & 0.173$^\dagger$  &   
\cite{Dutra:2012mb}   & \cr
SkT2-B & 0.970  & 242.000  & 33.500  & 58.000  &  & 0.186  & 0.174  & 
\cite{Brown:2013pwa}  & \cr
SkT2-T & 0.960  & 238.000  & 32.600  & 62.000  & -75.000  & 0.188  & 0.178  & 
\cite{Tsang:2019ymt,Dutra:2012mb}   & \cr
SkT3 & 1.000  & 235.740  & 31.500  & 55.310  & -132.050  & 0.182$^\dagger$  & 0.173$^\dagger$  &  
\cite{Dutra:2012mb}   & \cr
SkT3-B & 0.980  & 241.000  & 32.700  & 53.000  &  & 0.179  & 0.172  & 
\cite{Brown:2013pwa}  & \cr
SkT3-T & 0.970  & 236.000  & 31.900  & 58.000  & -80.000  & 0.183  & 0.178  & 
\cite{Tsang:2019ymt}  & \cr
Sk-T4 & 1.000  & 235.560  & 35.457  & 94.100  & -24.500  & 0.253  & 0.212$^\dagger $ & 
\cite{Steiner:2004fi,Inakura:2015cla,RocaMaza:2011pm}  & \cr
Sk-T6 & 1.000  & 235.950  & 29.970  & 30.900  & -211.530  & 0.151  & 0.155$^\dagger $ & 
\cite{RocaMaza:2011pm,Dutra:2012mb}   & \cr
Skxs20 & 0.960  & 201.950  & 35.500  & 67.060  & -122.310  & 0.201$^\dagger$  & 0.183$^\dagger$  & \cite{Dutra:2012mb}   & \cr
Skz2 & 0.700  & 230.070  & 32.010  & 16.810  & -259.660  & 0.120$^\dagger$  & 0.138$^\dagger$  &  \cite{Dutra:2012mb,Wang:2014rva}   & \cr
Skz4 & 0.700  & 230.080  & 32.010  & 5.750  & -240.860  & 0.102$^\dagger$  & 0.128$^\dagger$  &   \cite{Dutra:2012mb,Wang:2014rva}   & \cr
SLy0 & 0.700  & 229.670  & 31.982  & 44.873$^\dagger$  & -116.230  & 0.165  & 0.163$^\dagger$  & 
\cite{Steiner:2004fi,Dutra:2012mb}   & \cr
SLy10 & 0.680  & 229.740  & 31.980  & 38.740  & -142.190  & 0.155$^\dagger$  & 0.158$^\dagger$  & 
\cite{Ducoin:2010as,Dutra:2012mb}   & \cr
Sly2 & 0.700  & 229.920  & 32.000  & 47.460  & -115.130  & 0.170$^\dagger$  & 0.166$^\dagger$ & \cite{Dutra:2012mb,Fortin:2016hny}   & \cr
SLy230a & 0.700  & 229.890  & 31.980  & 44.310  & -98.210  & 0.155  & 0.158$^\dagger$  & 
\cite{Steiner:2004fi,Ducoin:2010as,Fortin:2016hny,Dutra:2012mb}    & \cr
Sly230b & 0.690  & 229.960  & 32.010  & 45.960  & -119.720  & 0.167$^\dagger$  & 0.164$^\dagger$  & \cite{Ducoin:2010as,Dutra:2012mb}   & \cr
SLy4 & 0.690  & 229.900  & 32.000  & 45.900  & -119.700  & 0.162  & 0.152  & 
\cite{RocaMaza:2011pm,Lim:2013tqa,Inakura:2015cla,Gonzalez-Boquera:2017rzy,Fortin:2016hny,Zhao:2016ujh}   & \cr
SLy4-B & 0.700  & 224.000  & 34.100  & 56.000  &  & 0.184  & 0.174  & 
\cite{Brown:2013pwa}  & \cr
SLy4-T & 0.760  & 222.000  & 33.600  & 66.000  & -55.000  & 0.191  & 0.179  & 
\cite{Tsang:2019ymt,Dutra:2012mb}   & \cr
SLy5 & 0.700  & 229.920  & 32.010  & 48.150  & -112.760  & 0.162  & 0.160  &
\cite{Dutra:2012mb,Zhao:2016ujh}   & \cr
SLy6 & 0.690  & 229.860  & 31.960  & 47.450  & -112.710  & 0.161  & 0.152   &
\cite{Dutra:2012mb,Zhao:2016ujh}   & \cr
Sly9 & 0.670  & 229.840  & 31.980  & 54.860  & -81.420  & 0.182$^\dagger$  & 0.172$^\dagger$  & \cite{Dutra:2012mb,Fortin:2016hny}   & \cr
SQMC650 & 0.780  & 218.110  & 33.650  & 52.920  & -173.150  & 0.178$^\dagger$  & 0.171$^\dagger$  & \cite{Dutra:2012mb}   & \cr
SQMC700 & 0.760  & 222.200  & 33.470  & 59.060  & -140.840  & 0.188$^\dagger$  & 0.176$^\dagger$  & \cite{Dutra:2012mb}   & \cr
SQMC750-B & 0.710  & 228.000  & 34.800  & 59.000  &  & 0.190  & 0.176  & 
\cite{Brown:2013pwa}  & \cr
SQMC750-T & 0.750  & 223.000  & 33.900  & 68.000  & -50.000  & 0.194  & 0.180  & 
\cite{Tsang:2019ymt,Dutra:2012mb}   & \cr
SR1 & 0.900  & 202.150  & 29.000  & 41.245$^\dagger$  &  & 0.160  & 0.160$^\dagger$  & 
\cite{Steiner:2004fi}  & \cr
SR2 &  & 224.640  & 30.071  & 49.130$^\dagger$  &  & 0.172  & 0.167$^\dagger$  & 
\cite{Steiner:2004fi}   & \cr
SR3 &  & 222.550  & 29.001  & 48.308$^\dagger$  &  & 0.171  & 0.166$^\dagger$  & 
\cite{Steiner:2004fi}   & \cr
SV & 0.380  & 306.000  & 32.800  & 96.100  & 24.190  & 0.230  & 0.196  & 
\cite{Lim:2013tqa,Ducoin:2010as,Zhao:2016ujh,Dutra:2012mb}    & \cr
SV-bas & 0.900  & 221.760  & 30.000  & 32.000  & -156.570  & 0.155  & 0.158$^\dagger$  & 
\cite{Reinhard:2016sce,Dutra:2012mb}  & \cr
SV-K218 & 0.900  & 218.230  & 30.000  & 35.000  & -206.870  & 0.161  & 0.161$^\dagger$  & 
\cite{Reinhard:2016sce,Dutra:2012mb}  & \cr
SV-K226 & 0.900  & 225.820  & 30.000  & 34.000  & -211.920  & 0.159  & 0.160$^\dagger$  & 
\cite{Reinhard:2016sce,Dutra:2012mb}   & \cr
SV-K241 & 0.900  & 241.070  & 30.000  & 31.000  & -230.770  & 0.151  & 0.155$^\dagger$  & 
\cite{Reinhard:2016sce,Dutra:2012mb}  & \cr
SV-kap00 & 0.900  & 233.440  & 30.000  & 40.000  & -161.780  & 0.158  & 0.159$^\dagger$  & 
\cite{Reinhard:2016sce,Dutra:2012mb}  & \cr
SV-kap20 & 0.900  & 233.440  & 30.000  & 36.000  & -193.190  & 0.155  & 0.158$^\dagger$  & 
\cite{Reinhard:2016sce,Dutra:2012mb}  & \cr
SV-kap60 & 0.900  & 233.450  & 30.000  & 29.000  & -249.750  & 0.154  & 0.157$^\dagger$  & 
\cite{Reinhard:2016sce,Dutra:2012mb}   & \cr
SV-L25 & 0.900  &  & 30.000  & 25.000  &  & 0.143  & 0.151$^\dagger$  & 
\cite{Reinhard:2016sce}   & \cr
SV-L32 & 0.900  &  & 30.000  & 32.000  &  & 0.154  & 0.157$^\dagger$  & 
\cite{Reinhard:2016sce}  & \cr
SV-L40 & 0.900  & 233.3 & 30.000  & 40.000  &  & 0.166  & 0.164$^\dagger$  & 
\cite{Reinhard:2016sce}  & \cr
SV-L47 & 0.900  & 233.4 & 30.000  & 47.000  &  & 0.177  & 0.170$^\dagger$  & 
\cite{Reinhard:2016sce}  & \cr
\hline
\end{tabular}
\end{table}

\setcounter{table}{0}
\begin{table}[h]
\caption{Properties of 206 EoSs (4).
}
\begin{tabular}{c|ccccccc|cc}
\hline
 & m*/m & K &J & L & Ksym & Rskin-208 & Rskin-48 & Refs. \cr
\hline
SV-mas07 & 0.700  & 233.540  & 30.000  & 52.000  & -98.770  & 0.152  & 0.156$^\dagger$  & 
\cite{Reinhard:2016sce,Dutra:2012mb}  & \cr
SV-mas08 & 0.800  & 233.130  & 30.000  & 40.000  & -172.380  & 0.160  & 0.160$^\dagger$  & 
\cite{Reinhard:2016sce,Dutra:2012mb,Wang:2014rva}  & \cr
SV-mas10 & 1.000  & 234.330  & 30.000  & 28.000  & -252.500  & 0.152  & 0.156$^\dagger$  & 
\cite{Reinhard:2016sce,Dutra:2012mb}  & \cr
SV-sym28 & 0.900  & 240.860  & 28.000  & 7.000  & -305.940  & 0.117  & 0.136$^\dagger$  & 
\cite{Reinhard:2016sce,Dutra:2012mb}  & \cr
SV-sym32 & 0.900  & 233.810  & 32.000  & 57.000  & -148.790  & 0.192  & 0.178$^\dagger$  & 
\cite{Reinhard:2016sce,Dutra:2012mb}  & \cr
SV-sym32-B & 0.910  & 237.000  & 32.300  & 51.000  &  & 0.176  & 0.174  & 
\cite{Brown:2013pwa}  & \cr
SV-sym32-T & 0.910  & 232.000  & 31.500  & 58.000  & -77.000  & 0.181  & 0.179  & 
\cite{Tsang:2019ymt}  & \cr
SV-sym34 & 0.900  & 234.070  & 34.000  & 81.000  & -79.080  & 0.227  & 0.198$^\dagger$  &   
\cite{Reinhard:2016sce,Dutra:2012mb,Wang:2014rva}  & \cr
TFa &  & 245.100  & 35.050  & 82.500  & -68.400  & 0.250  & 0.210$^\dagger$  & 
\cite{Fattoyev:2013yaa} & \cr
TFb &  & 250.100  & 40.070  & 122.500  & 45.800  & 0.300  & 0.238$^\dagger$  & 
\cite{Fattoyev:2013yaa}  & \cr
TFc &  & 260.500  & 43.670  & 135.200  & 51.600  & 0.330  & 0.255$^\dagger$  & 
\cite{Fattoyev:2013yaa} & \cr
TM1 &  & 281.000  & 36.900  & 110.800  & 33.550  & 0.272$^\dagger$  & 0.223$^\dagger$  & 
\cite{Ishizuka:2014jsa,Ducoin:2010as,Fortin:2016hny,Chen:2010qx}  & \cr
TW99 &  & 241.000  & 32.800  & 55.000  & -124.000  & 0.196  & 0.186  &  
\cite{Chen:2010qx,Zhao:2016ujh}  &\cr 
UNEDF0 &  & 229.800  & 30.500  & 45.100  & -189.600  & 0.166$^\dagger$  & 0.164$^\dagger$  & 
\cite{Inakura:2015cla,Dutra:2012mb} &\cr 
UNEDF1 &  & 219.800  & 29.000  & 40.000  & -179.400  & 0.158$^\dagger$  & 0.159$^\dagger$  & 
\cite{Inakura:2015cla}  & \cr
Z271 &  & 271.000  & 35.369  & 89.520$^\dagger$  &  & 0.238  & 0.204$^\dagger$  & 
\cite{Steiner:2004fi}   & \cr
\hline
\end{tabular}
\end{table}   


\squeezetable
\begin{table}[h]
\caption{
Parameter sets of D1MK and D1PK. 
}
 \begin{tabular}{c|c|cccc|ccc|c}
\hline
  D1MK &$\mu_i$&$W_i$ & $B_i$ & $H_i$ & $M_i$ &
  $t_0^i$ & $x_0^i$ & $\alpha_i$& $W_0$\cr
\hline
   $i=1$ & 0.5 &  -17242.0144 & 19604.4056 & -20699.9856& 16408.6002&
  1561.7167 & 1 & 1/3  & 115.36 \cr
  $i=2$ & 1.0 & 642.607965& -941.150253& 865.572486& -845.300794&
  0 & -1 & 1 & \cr
\hline
  D1PK &$\mu_i$&$W_i$ & $B_i$ & $H_i$ & $M_i$ &
  $t_0^i$ & $x_0^i$ & $\alpha_i$& $W_0$\cr
\hline
   $i=1$ & 0.90 &  -465.027582 & 155.134492 & -506.775323& 117.749903&
  981.065351& 1 & 1/3  & 130\cr
  $i=2$ & 1.44 &34.6200000& -14.0800000& 70.9500000& -41.3518104&
  534.155654 & -1 & 1 & \cr
\hline
\end{tabular}
\end{table}  

\end{document}